\documentclass[12pt]{article}
 \usepackage {graphicx}
 \usepackage[latin1]{inputenc}
 \usepackage{amsmath}
 \usepackage{amsfonts}
 \usepackage{amssymb}
 \usepackage{amsthm}
      \usepackage{mathrsfs}
   \usepackage{kformat}

\newcommand{\bfom}{\mbox{\boldmath $\omega$}}
\newcommand{\bfPhi}{\mbox{\boldmath $\Phi$}}
\newcommand{\cc}{\textrm{c}}
\newcommand{\s}{\textrm{s}}
\newcommand{\bfb}{\hat{\textbf{b}}}

\newcommand{\bfs}{\hat{\textbf{s}}}
\newcommand{\bfp}{\textbf{p}}
\newcommand{\bfq}{\textbf{q}}
\newcommand{\bfQ}{\textbf{Q}}
\newcommand{\bfP}{\textbf{P}}
\newcommand{\bfG}{{\textbf{G}}}
\newcommand{\bfi}{\hat{\textbf{i}}}
\newcommand{\bfl}{\hat{\textbf{l}}}
\newcommand{\bfj}{\hat{\textbf{j}}}
\newcommand{\bfv}{\textbf{v}}
\newcommand{\bfu}{\textbf{u}}

\newcommand{\bfg}{\textbf{g}}

\newcommand{\inc}{{\it i}}

 \newcommand{\be}{\begin{equation}}
 \newcommand{\ee}{\end{equation}}
 \newcommand{\ba}{\begin{eqnarray}}
 \newcommand{\ea}{\end{eqnarray}}
 \newcommand{\efbold}{\mbox{{\boldmath $f$}}}

 \newcommand{\erbold}{\mbox{{\boldmath $r$}}}
 \newcommand{\Phibold}{\mbox{{\boldmath $\Phi$}}}
 
 \newcommand{\mubold}{\mbox{{\boldmath $\mu$}}}

   \newcommand{\Omegabold}{\mbox{{\boldmath $ \Omega $}}}
    \newcommand{\omegabold}{\mbox{{\boldmath $ \omega $}}}

 \newcommand{\doterbold}{\dot{\textbf {\mbox{\boldmath ${\boldmath r}$}} }}

\newcommand{\Eqref}[1]{(\ref{#1})}
\newcommand{\Frac}[2]{\mbox{$\displaystyle\frac{#1}{#2}$}}
\newcommand{\Int}{\displaystyle\int}
\newcommand{\Sqrt}[1]{\displaystyle\sqrt{#1}}

\def\am{\mathop{\rm am}\nolimits}

\title{
 The Serret-Andoyer Formalism in Rigid-Body Dynamics: I. Symmetries and Perturbations}

 \author{Pini Gurfil\\
 {\it Faculty of Aerospace Engineering} \\{\it Technion - Israel
 Institute of Technology, Haifa 32000 Israel}\\
 ~\\
 Antonio Elipe\\
 {\it Grupo de Mecanica Espacial} \\{\it Universidad de Zaragoza, Zaragoza, Spain}\\
 ~\\
 William Tangren\\
 {\it US Naval Observatory, Washington DC 20392 USA}\\
 ~\\
 and\\
 ~\\
 Michael Efroimsky\\
 {\it US Naval Observatory, Washington DC 20392 USA}\\
 }

\date{}
\begin{document}

\maketitle

\begin{abstract}
 This paper reviews the Serret-Andoyer (SA) canonical formalism in
 rigid-body dynamics, and presents some new results. As is well known,
 the problem of unsupported and unperturbed rigid rotator can be
 reduced. The availability of this reduction is offered by the
 underlying symmetry, that stems from conservation of the angular
 momentum and rotational kinetic energy. When a perturbation is
 turned on, these quantities are no longer preserved. Nonetheless,
 the language of reduced description remains extremely instrumental
 even in the perturbed case. 
%
 We describe the canonical reduction performed by the Serret-Andoyer (SA)
 method, and discuss its applications to attitude dynamics and to the theory
 of planetary rotation. Specifically, we consider the case of
 angular-velocity-dependent torques, and discuss the 
 variation-of-parameters-inherent antinomy between canonicity and osculation. Finally, we address the
 transformation of the Andoyer variables into action-angle ones, using
 the method of Sadov.
\end{abstract}


\section{INTRODUCTION}

One of the classical problems of mechanics is that of a free motion of a rigid
body, usually referred to as the \emph{Euler-Poinsot} problem. The formulation and
solution of this problem are usually performed in two steps. First, the
\emph{dynamics} of the rotation are represented by differential equations for the
components of the body angular velocity. As is well known, these equations admit a
closed-form solution in terms of Jacobi's elliptic functions \cite{marsden}.
Second, the \emph{kinematic} equations are utilised to transform the body angular
velocity into a spatial inertial frame.  While this classical formulation is
widespread among engineers, astronomers exhibit a marked preference for formulation
that takes advantage of the internal symmetries of the Euler-Poinsot setting
\cite{depritelipe}. The existence of internal symmetries indicates that the
unperturbed Euler-Poinsot problem can be reduced to a smaller number of variables,
whereafter the disturbed setting can be treated as a perturbation expressed through
those new variables (the disturbance may be called into being by various reasons -
physical torques; inertial torques emerging in non-inertial frames of reference;
non-rigidity of the rotator).

There exist reasons for performing this reduction in the Hamiltonian
form and making the resulting reduced variables canonical. The first
reason is that a perturbed Hamiltonian system can be analytically
solved in any order over the parameter entering the
perturbation.\footnote{~While in the past such analytical solutions
were built by means of the von Zeipel method, more efficient is the
procedure independently offered by Hori \cite{Hori} and Deprit
\cite{deprit_1969}. An explanation of the method can be found in
\cite{Kholshevnikov_1973}. A concise introduction into the subject
can be found in Chapter 8 of the second volume of
\cite{Boccaletti_and_Pucacco_2002} and in Chapter 5 of
\cite{Kholshevnikov_1985}.} A celebrated example of a successful
application of the Hori-Deprit method to a geophysical problem is
given by the theory of rigid Earth rotation offered by Kinoshita
\cite{kinoshita77} and further developed by joint efforts of Escapa,
Getino, \& Ferr{\'a}ndiz \cite{escapa_2001, escapa_2002}, Getino \&
Ferr{\'a}ndiz \cite{getino}, and Kinoshita \& Souchay
\cite{kinoshita_souchay}. Other applications to planetary and liunar
rotation may be found in \cite{moons82}, \cite{navarrof2}, and
\cite{henrard}.

The second advantage of the canonical description originates from
the convenience of numerical implementation: symplectic schemes are
well-known for their good stability and precision. (See, for
example, Yoshida \cite{yoshida}.) This is why the Hamiltonian
methods are especially beneficial at long time scales, a feature
important in astronomy. (Laskar \& Robutel \cite{laskar}; Touma \&
Wisdom \cite{touma93,touma})

 Several slightly different sets of canonical variables are used
 for modeling rigid-body dynamics and kinematics. Most popular is
 the set suggested in 1923 by Andoyer \cite{andoyer}. This set is not
 completely reduced: while three of its elements are constants (in
 the unperturbed free-spin case), three others are permitted to
 evolve. Andoyer arrived at his variables through a pretty
 sophisticated procedure. Canonicity of Andoyer's change
 of variables is not immediately apparent (though can be proven by a
 direct construction of the corresponding generating function). Much
 later, the study by Andoyer was amended by Deprit \cite{deprit} who
 demonstrated that the canonicity of Andoyer's transformation may be
 proven by using differential forms and without resorting to explicitly
 finding a generating function.

A full reduction of the Euler-Poinsot problem, one that has the
Andoyer variables as its starting point, was recently offered by
Deprit \& Elipe \cite{depritelipe}. It would be interesting to
notice that, historically, the pioneer canonical treatment of the
problem, too, was formulated in a completely reduced way, i.e., in
terms of canonically-conjugate constants of motion. These
constants, presently known as Serret elements, are named after the
19th century French mathematician Joseph Alfred Serret who
discovered these variables by solving the Jacobi equation
written in terms of the Eulerian coordinates \cite{serret}.
Serret's treatment was later simplified by Radeau \cite{radeau}
and Tisserand \cite{tisserand}. However, for the first time the
Serret elements appeared in an earlier publication by Richelot
\cite{richelot}.

Probably the most distinguishing feature of the canonical approach
is that it permits reduction of the torque-free rotational dynamics
to one-and-a-half degrees of freedom. In essence, such a
formulation reduces the dynamics by capturing the underlying
symmetry of the free rigid body problem -- the symmetry taking its
origin from the conservation of energy and angular momentum. As a
result, the entire dynamics can be expressed by differential
equations for two of the Eulerian angles and one of the conjugate
momenta -- equations that are readily integrable by quadrature. The
corresponding one-and-a-half-degrees-of-freedom Hamiltonian yields a
phase portrait, that is similar to that of the simple pendulum and
contains a separatrix confining the librational motions
\cite{deprit}.

    In the canonical formulation of attitude-mechanics problems,
    incorporation of perturbation and/or control torques into the
    picture is a subtle operation.
In particular, when the perturbing torque is
angular-velocity-dependent, the canonicity demand contradicts the osculation condition. In other words, the
expression for the angular velocity through the canonical Serret
or Andoyer variables acquires a correction called the
\emph{convective} term.

To better understand the latter observation, we shall use an
orbital-dynamics analogy. This shall be convenient, because both
orbital and rotational dynamics employ the method of
variation-of-parameters to model perturbing inputs.  In both
cases, the coordinates (Cartesian, in the orbital case, or
Eulerian in the rotational case) are expressed, in a non-perturbed
setting, via the time and six adjustable constants called elements
(orbital elements or rotational elements, accordingly). If, under
disturbance, we use these expressions as an ansatz and endow the
``constants" with time dependence, then the perturbed velocity
(orbital or angular) will consist of a partial derivative with
respect to the time, plus the convective term, one that includes
the time derivatives of the variable ``constants." Out of sheer
convenience, the so-called Lagrange constraint is often imposed.
This constraint nullifies the convective term and, thereby,
guarantees that the functional dependence of the velocity upon the
time and ``constants" stays, under perturbations, the same as it
used to be in the undisturbed setting. The variable ``constants"
obeying this condition are called osculating elements. Otherwise,
they are simply called orbital elements (in orbital mechanics) or
rotational elements (in attitude mechanics).

When the dynamical equations, written in terms of the
``constants," are demanded to be canonical, these ``constants" are
the Delaunay elements in the orbital case, or the initial values
of the Andoyer elements, in the spin case. These two sets of
elements share a feature not readily apparent: in certain cases,
the standard equations render these elements non-osculating.


In attitude dynamics, the Andoyer variables come out
non-osculating when the perturbation depends upon the angular
velocity. For example, since a transition to a non-inertial frame
is an angular-velocity-dependent perturbation, then amendment of
the dynamical equations by only adding extra terms to the
Hamiltonian makes these equations render non-osculating Andoyer
variables. To make them osculating, extra terms must be added in
the dynamical equations (and then these equations will no longer
be symplectic). Calculations in terms of non-osculating variables
are mathematically valid, but their physical interpretation is not
always easy.

The purpose of this paper will be threefold. We shall first aim to
provide the reader with a coherent review of the Serret-Andoyer (SA)
formalism; second, we shall dwell on the canonical perturbation
theory in the SA context; and third, we shall consider introduction
and physical interpretation of the Andoyer variables in precessing
reference frames, with applications to planetary rotation. A
subsequent publication by Bloch, Gurfil, \& Lum (2007) \cite{paper2}
will deal with the Andoyer-variables-geometry based on the theory of
Hamiltonian systems on Lie groups, and will fill the existing gap in
the control literature by using the SA modeling of rigid-body
dynamics in order to derive nonlinear asymptotically stabilising
controllers.

\section{THE EULERIAN VARIABLES}
\label{S:2}
Through this entire section the compact notations
$\,s_x=\,\sin(x)\,,\,\;c_x=\,\cos(x)\;$ will be used. The
angular-velocity and angular-momentum vectors, as seen in the
principal axes of the body, will be denoted by low-case letters:
$\;{{\omegabold}}\;$ and $\;{\bf{g}}\;$. The same two vectors in an
inertial coordinate system will be denoted with capital letters:
$\;{{\Omegabold}}\;$ and $\;{\bf{G}}\;$.

\subsection{Basic formulae}

Consider a rotation of a rigid body about its center of mass, $O$.
The body frame, $
B$, is a Cartesian dextral frame that is
centered at the point $O$ and is defined by the unit vectors
$\bfb_1,\,\bfb_2$, constituting the fundamental plane, and by
$\bfb_3=\bfb_1\times\bfb_2$. The attitude of $
 B$ will be
studied relative to an inertial Cartesian dextral frame, $
I$, defined by the unit vectors $\bfs_1$, $\bfs_2$ lying on the
fundamental plane, and by $\bfs_3=\bfs_1\times\bfs_2$. These
frames are depicted in Fig.~\ref{frames}.

A transformation from $
I$ to $
 B$ may be
implemented by three consecutive rotations making a $3-1-3$
sequence. Let the line of nodes (LON) $OQ_2$ be the intersection
of the body fundamental plane and the inertial fundamental plane,
as shown in Fig.~\ref{frames}. Let $\bfl$ be a unit vector
pointing along $OQ_2$. The rotation sequence can now be defined as
follows:\footnote{~Be mindful that in the physics and engineering
literature the Euler angles are traditionally denoted with
$\,(\,\phi\,,\,\theta\,,\,\psi\,)\,$. In the literature on the
Earth rotation, the inverse convention,
$\,(\,\psi\,,\,\theta\,,\,\phi\,)\,$, is in use.
\cite{kinoshita77, escapa_2001, escapa_2002, getino,
kinoshita_souchay}}

\begin{itemize}

\item $R(\phi,\bfs_3)$, a rotation about $\bfs_3$ by
$0\leq\phi\leq 2\pi$, mapping $\bfs_1$ onto $\bfl\;$;

\item $R(\theta,\bfl)$, a rotation about $\bfl$ by
$0\leq\theta\leq \pi$, mapping $\bfs_3$ onto $\bfb_3\;$;

\item $R(\psi,\bfb_3)$, a rotation about $\bfb_3$ by
$0\leq\psi\leq 2\pi$, mapping $\bfl$ onto $\bfb_1\;$.

\end{itemize}

The composite rotation, $R\in SO(3)$, transforming any inertial
vector into the body frame, is given by
 \ba
 R(\phi,\theta,\psi) =R(\psi,\bfb_3)R(\theta,\bfl)R(\phi,\bfs_3)
 \label{R}
 \label{1}
 \ea
Evaluation of this product gives
 \begin{equation}
 \label{Rmat}
    R(\phi,\theta,\psi)=\left[%
\begin{array}{ccc}
 \cc_\psi\cc_\phi-\s_\psi\cc_\theta\s_\phi & \cc_\psi\s_\phi+\s_\psi\cc_\theta\cc_\phi & \s_
 \psi\s_\theta \\
 -\s_\psi\cc_\phi-\cc_\psi\cc_\theta\s_\phi & -\s_\psi\s_\phi+\cc_\psi\cc_\theta\cc_\phi &
 \cc_\psi\s_\theta \\
 \s_\theta\s_\phi & -\s_\theta\cc_\phi & \cc_\theta \\
\end{array} %
\right].
 \end{equation}

To write the kinematic equations, we recall that the \emph{body
angular-velocity} vector, $\bfom=[\omega_1,\omega_2,\omega_3]^T$
satisfies \cite{kane}
 \ba
 \widehat{\bfom} = - \dot R R^T
 \label{3}
 \label{omegas}
 \ea
where the hat map $\widehat{(\cdot)} : \mathbb R^3 \to
\mathfrak{so}(3)$ is the usual Lie algebra isomorphism and
 \begin{equation}
  \widehat{\bfom} =\left[%
 \begin{array}{ccc}
  0 & -\omega_3 & \omega_2 \\
  \omega_3 & 0 & -\omega_1 \\
  -\omega_2 & \omega_1 & 0  \\
\end{array}%
\right].
  \label{4}
 \end{equation}
Insertion of Eq.~(\ref{Rmat}) into (\ref{omegas}) yields the
well-known expressions for the components of the vector of the
body angular velocity $\;\omegabold\;$:
\begin{eqnarray}
  \omega_1 &=& \dot\phi\s_\theta
  \s_\psi+\dot{\theta}\cc_\psi\;\;\;,
   \label{omega1}
   \label{5}\\
  \omega_2 &=&
  \dot{\phi}\cc_\psi\s_\theta-\dot{\theta}\s_\psi\;\;\;,
   \label{omega2}
   \label{6}\\
  \omega_3 &=& \dot{\psi}+\dot{\phi}\cc_\theta\;\;\;,
   \label{omega3}
   \label{7}
\end{eqnarray}
while the action of the rotation matrix (\ref{Rmat}) upon the body
angular velocity gives the inertial-frame-related (sometimes called
spatial) angular velocity:
 \ba
 \Omegabold\;=\;R^T\;\omegabold\;\;\;,
 \label{8}
 \ea
with the following components:
 \ba
 \Omega_1\;=\; \dot{\theta} \; c_\phi \;+\; \dot{\psi} \; s_\theta \;
 s_\phi\;\;\;,
 \label{9}
 \label{Omega1} \\
 \Omega_2\;=\; \dot{\theta} \; s_\phi \;+\; \dot{\psi} \; s_\theta \;
 c_\phi\;\;\;,
 \label{10}
 \label{Omega2}\\
 \Omega_3\;=\; \dot{\phi} \;+\; \dot{\psi} \;c_\theta
 ~~~.\,~~~~~~\,
 \label{11}
 \label{Omega3}
 \ea
In the body frame, attitude dynamics are usually formulated by means
of the \emph{Euler-Poinsot} equations. In a free-spin case, these
equations look as
 \begin{eqnarray}
    {\mathbb I}\dot\bfom+\bfom\times\mathbb I\bfom =  0\;\;\;,
 \label{12}
 \label{EPfree}
 \end{eqnarray}
 $\mathbb{I}\;$ being the inertia tensor.
To arrive to these equations, one has to start out with the
Lagrangian,
 \begin{eqnarray}
 \mathcal L(\phi,\theta,\psi,\dot\phi,\dot\theta,\dot\psi) = \frac{1}{2}\bfom\cdot{\mathbb I}
 \bfom\;\;\;,
 \label{13}
 \label{lag}
 \end{eqnarray}
to substitute therein expressions (\ref{omega1} - \ref{omega3});
to write down the appropriate Euler-Lagrange variational equations
for the Euler angles; and then to use the expressions
(\ref{omega1} - \ref{omega3}) again, in order to rewrite these
Euler-Lagrange equations in terms of $\;\omegabold\;$. This will
result in (\ref{EPfree}). The same sequence of operations carried
out on a perturbed Lagrangian $\;{\cal L}\,+\,{\Delta \cal L}\;$
will produce the forced Euler-Poinsot equations:
 \begin{eqnarray}
    {\mathbb I}\dot\bfom+\bfom\times{\mathbb I}\bfom =  \bfu\;\;\;,
 \label{14}
 \label{EP}
 \end{eqnarray}
 $\bfu\in\mathbb R^3\,
 $ being the body-frame-related torque that can be expressed through derivatives of $\,\Delta
 {\cal L}\,$. The Euler-Poinsot description of the motion is essentially Lagrangian.
 Alternatively, the equations of attitude dynamics can be cast into a Hamiltonian shape.

 A most trivial but important observation can be made simply from looking at (\ref{14}). If
 one rewrites (\ref{14}) not in terms of velocities but in terms of the Euler angles, three differential
 equations of the second order will emerge. Their solution will depend on the time and six adjustable
 constants. Hence, no matter which description one employs --
 Lagrangian or Hamiltonian -- the number of emerging integration constants will always be
 six.

 \subsection{Hamiltonian description. The free-spin case.}

 From here forth we shall assume that the body axes coincide with the principal axes
 of inertia, so we can write
 \ba
 \mathbb I = \textrm{diag}(I_1,I_2,I_3).
 \label{15}
 \label{I}
 \ea
Having substituted expressions (\ref{omega1} - \ref{omega3}) and
(\ref{I}) into (\ref{lag}), one can easily write the generalised
momenta $\;p_n\;=\;\Phi\,,\;\Theta\,,\;\Psi\;$ conjugate to the
configuration variables $\;q_n\;=\;\phi\,,\;\theta\,,\;\psi\;$
\cite{depritelipe,bloch}:

 \begin{eqnarray}
  \Phi &=& \frac{\partial \mathcal L}{\partial\dot\phi} =
  I_1\s_\theta\s_\psi(\dot{\phi}\s_\psi\s_{\theta}+
  \dot{\theta}\cc_\psi)+I_2\s_\theta\cc_\psi(\dot{\phi}\cc_\psi\s_\theta-\dot{\theta}\s_\psi)
  +I_3\cc_\theta(\dot{\phi}\cc{\theta}+\dot\psi)\;\;\;,~~~~~~~~
  \label{16}\\
  \Theta &=& \frac{\partial \mathcal L}{\partial\dot\theta} =
  I_1\cc_\psi(\dot{\phi}\s_\theta\s_\psi+\dot{\theta}\cc_\psi)-I_2\s_\psi(\dot{\phi}\cc_\phi
  \s_\theta - \dot{\theta}\s_\psi)\;\;\;,
  \label{17}\\
  \Psi &=& \frac{\partial \mathcal L}{\partial\dot\psi} =
  I_3(\dot{\phi}\cc_\theta+\dot{\psi})\;\;\;.
  \label{18}
 \end{eqnarray}
 \noindent
The inverse relations will assume the form of
 ~\\
\begin{eqnarray}
 \dot\phi &=& -\frac{I_1\cc_\psi(\Psi\cc_\theta \cc_\psi-\Phi\cc_\psi+\Theta\s_\theta\s_\psi)
 +I_2\s_\psi(\Psi\cc_\theta\s_\psi-\Phi\s_\psi-\Theta\cc_\psi\s_\theta)}{I_1 I_2\s_\theta^2}~
 ~~,
  \label{19}
  \label{phid}\\
  \nonumber\\
  \dot\theta &=& \frac{I_2\cc_\psi(\Phi\s_\psi+\Theta\s_\theta\cc_\psi-\Psi\s_\psi\cc_\theta)
  + I_1\s_\psi(\Psi\cc_\psi\cc_\theta-\Phi\cc_\psi+\Theta\s_\theta\s_\psi)}{I_1
  I_2\s_\theta}\;\;\;,
  \label{20}\\
  \nonumber\\
 \dot\psi &=&-\frac{I_1I_3\cc_\psi(\Phi\cc_\theta\cc_\psi-\Psi\cc_\psi\cc_\theta^2-\Theta
 \s_\psi \s_\theta\cc_\theta)
  +I_3I_2\s_\psi(\Phi\cc_\theta\s_\psi-\Psi\cc_\theta^2\s_\psi+\Theta\cc_\psi\s_\theta
  \cc_\theta)}{I_1I_2I_3\s_\theta^2}\nonumber\\&+&\frac{\Psi}{I_3}
  ~~~~~~~~
  \label{21}
  \label{psid}
\end{eqnarray}
Substitution of (\ref{phid} - \ref{psid}) into the
Legendre-transformation formula
 \ba
 \label{22}
 \label{legendre}
 {\mathcal H}\;=\;\Phi\;\dot\phi\;+\;\Theta\;\dot\theta\;+\;\Psi\;\dot\psi\;-\;{\mathcal L}
 \ea
will then readily give us the free-spin Hamiltonian:
 \begin{eqnarray}
 \nonumber
 {\mathcal H}(\phi,\theta,\psi,\Phi,\Theta,\Psi)&=&\frac{1}{2}\left(\frac{\s_\psi^2}{I_1}+
 \frac{\cc_\psi^2}{I_2}\right)\left(\frac{\Phi-\Psi\cc_\theta}{\s_\theta}\right)^2+
 \frac{\Psi^2}{2I_3}+\frac{1}{2}\left(\frac{\cc_\psi^2}{I_1}+\frac{\s_\psi^2}{I_2}\right)
 \Theta^2 \\
     \nonumber\\
     \nonumber\\
    &+&\left(\frac{1}{I_1}-\frac{1}{I_2}\right)\left(\frac{\Phi-\Psi\;\cc_\theta}{\s_\theta}
    \right)
    \Theta\;\s_\psi\;\cc_\psi.
     \label{23}
     \label{hamil}
 \end{eqnarray}
In the above Hamiltonian the coordinate $\phi$ (the angle of
rotation about the inertial axis $\;{\bf{s}}_3\;$) is cyclic
(ignorable), so that the appropriate momentum $\;\Phi\;$ is an
integral of motion. This symmetry implies a possibility of
reduction of the free-spin problem to only two degrees of freedom.
Based on this observation, Serret \cite{serret} raised the
following question: Is there a canonical transformation capable of
reducing the number of degrees of freedom even further? We shall
discuss this issue in the following section. To get this
program accomplished, we shall need to know the relationship
between the body rotational angular momentum and the conjugate
 momenta. Let ${\bf g}= \sum g_i{\bf b}_i
 $ and ${\bf
 G}=\sum G_i{\bf s}_i
 $ be the angular
momentum in the body frame and in inertial frame, accordingly. By
plugging (\ref{omega1} - \ref{omega3}) into
%
 \begin{equation}
    \bf g={\mathbb I}\bfom
     \label{24}
 \end{equation}
and by subsequent insertion of (\ref{phid} - \ref{psid}) therein,
one easily arrives at
 \begin{eqnarray}
  g_1 &=& \frac{\Phi\s_\psi+\Theta\s_\theta\cc_\psi-\Psi\s_\psi\cc_\theta}{\s_\theta}
   \label{25}
   \label{g1} \;\;\;,\\
  g_2 &=&
  \frac{\Phi\cc_\psi-\Theta\s_\psi\s_\theta-\Psi\cc_\psi\cc_\theta}{\s_\theta}~~~,
  \label{26}
   \\
  g_3 &=& \Psi ~~~.
   \label{27}
   \label{g3}
 \end{eqnarray}
Notice the symplectic structure defined by the components of the angular
momentum. It is a matter of computing partial derivatives to check that the
Poisson brackets are
\begin{equation}\label{symplecticRB}
(g_1,g_2) = -g_3, \qquad (g_2,g_3) = -g_1, \qquad  (g_3,g_1) = -g_2
\end{equation}
 By the same token, insertion of expressions (\ref{Omega1} - \ref{Omega3}) into
  \ba
    {\bf G}= {\mathbb I}\Omegabold \;\;\;,
  \label{28}
 \ea
with the subsequent use of (\ref{phid} - \ref{psid}), entails:
\begin{eqnarray}
  G_1 &=&
  \frac{\Psi\s_\phi+\Theta\s_\theta\cc_\phi-\Phi\s_\phi\cc_\theta}{\s_\theta}~~~,
  \label{29}
   \label{G1}\\
  G_2 &=&
  \frac{\Phi\cc_\theta\cc_\phi-\Psi\cc_\phi+\Theta\s_\phi\s_\theta}{\s_\theta}~~~,
  \label{30}
   \\
  G_3 &=& \Phi~~~,
   \label{31}
   \label{G3}
\end{eqnarray}
yielding the symplectic structure

\begin{equation}\label{}
(G_1,G_2) = G_3, \qquad (G_2,G_3) = G_1, \qquad  (G_3,G_1) = G_2.
\end{equation}
from the symplectic structure \Eqref{symplecticRB} and the
expression \Eqref{hamil} of the Hamiltonian, there results
\begin{eqnarray}
 \dot{g}_1 &=& (g_1,\mathcal{H}) =
    - \left(\Frac{1}{I_2} - \Frac{1}{I_3}\right) g_2 g_3 ,\label{firstg1}\\[2ex]
 \dot{g}_2  &=& (g_2,\mathcal{H}) =
    -  \left(\Frac{1}{I_3} - \Frac{1}{I_1} \right) g_3 g_1 ,\\[2ex]
 \dot{g}_1  &=& (g_2,\mathcal{H}) =
    -  \left(\Frac{1}{I_1} - \Frac{1}{I_1} \right) g_1 g_2
    .\label{firstg3}
\end{eqnarray}

\begin{center}
 \begin{figure}
 \includegraphics[width=5in]{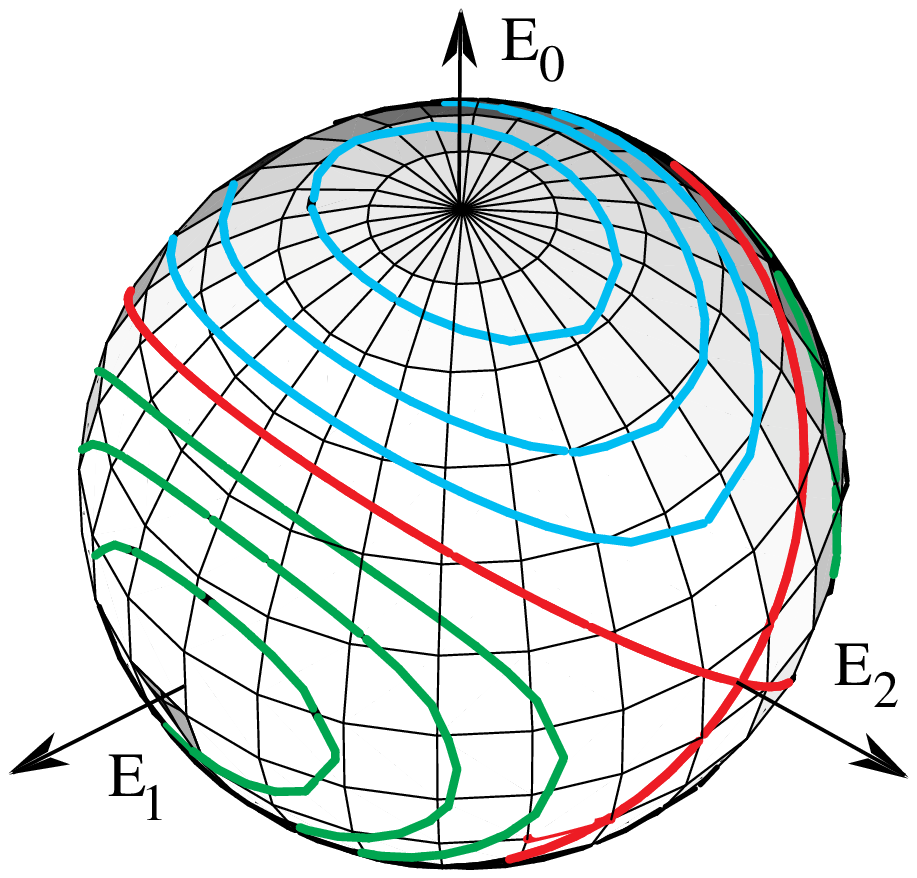}
 \caption{Phase flow of the Euler-Poinsot problem}
 \label{fi:sphere}
 \end{figure}
 \end{center}

This system is integrable because it admits two integrals, the
energy \Eqref{lag} and the norm of the angular momentum $\,|{\bf{G}}
|\, = G $. With this integral, we may regard the phase space of
(\ref{firstg1})-(\ref{firstg3}) as a foliation of invariants
manifolds
\begin{equation}\label{sphere}
S^2(G) = \{ (g_1, g_2, g_3) | g_1^2 +  g_2^2 + g_3^2 =  G^2 \}.
\end{equation}
The trajectories will be the level contours of the \emph{energy
ellipsoid},
 \begin{equation}
 \label{kinetic}
 T_{kin} = \Frac{1}{2} \;{\omegabold} \cdot {\bf{G}} =
  \Frac{1}{2}\left( I_1 \omega_1^2 + I_2 \omega_2^2 +I_3 \omega_3^2 \right)
  =  \Frac{1}{2}\left(\Frac{ g_1^2 }{I_1}+ \Frac{ g_2^2 }{I_2} +
  \Frac{ g_3^2 }{I_3} \right),
\end{equation}
on the sphere \Eqref{sphere}, as can be seen in Figure
\ref{fi:sphere}.


Below we shall need also the expressions for the canonical momenta
via the components of the angular momentum $\bf G$. As agreed above,
the Euler angles $\phi,\,\theta,\,\psi$ determine the orientation of
the body relative to some inertial reference frame. Let now the
angles $\phi_o,\,J,\,l$ define the orientation of the body relative
to the invariable plane (one orthogonal to the angular-momentum
vector $\bf G$), as in Fig.~\ref{frames}; and let
$h,\,I,\,g-\phi_o\,$ be the Euler angles defining the orientation of
the invariable plane relative to the reference one.\footnote{~It
would be natural to denote the orientation of the body relative to
the invariable plane with $\;\phi_o\,,\;\theta_o\,,\;\psi_o\;$, but
we prefer to follow the already established notations.} Evidently,
 \ba
 g_1\;=\;G\;s_{J}\;s_{l}
 \;\;\;,\;\;\;\;\;
 g_2\;=\;G\;s_{J}\;c_{l}
 \;\;\;,\;\;\;\;\;
 g_3\;=\;G\;c_{J}\;\;\;.
 \label{32a}
 \ea
It is now straightforward from (\ref{25} - \ref{27}) and (\ref{32a})
that
 \ba
 \Phi\;=\;g_1\,s_{\theta}\,s_{\psi}\,+\,g_2\,s_{\theta}\,c_{\psi}\,+\,g_3\,c_{\theta}\;=\;
 G\;\left(\,c_\theta \,c_{J}\,+\,s_{\theta}\,s_{J}\,c_{(\psi-l)}\,\right)
 \;=\;G\;c_I\;\;\;,\;\;\;
 \label{33}
 \ea
 \ba
 \Theta\,=\,g_1\,c_\psi\,-\;g_2\,s_\psi\;=\;G\;s_{J}\;s_{(l\,-\,\psi)}
 \;\;\;,\;\;\;~~~~~~~~~~~~~~~~~~~~~~~~~~~~~~~~~~~~~~~~~~~~~
 \ea
 \ba
 \Psi\;=\;g_3\;=\;G\;c_{J}\;\;\;.\;\;\;\;\;\;~~~~~~~~~~~~~~~~~~~~~~~~~~~~~~~~~~~~~~
 ~~~~~~~~~~~~~~~~~~~~~~~~~~~~
 \label{35}
 \ea

 \section{THE SERRET-ANDOYER\\ TRANSFORMATION}

 \subsection{Richelot (1850), Serret (1866), Radau (1869),\\ Tisserand (1889)}

 The method presently referred to as the Hamilton-Jacobi one is based on
 the Jacobi equation derived circa 1840. Though Jacobi's book \cite{jacobi}
 was published only in 1866, the equation became known to the scientific
 community already in 1842 when Jacobi included it into his lecture course.
 In 1850 Richelot \cite{richelot} suggested six constants of motion grouped
 into three canonically-conjugate pairs. These constants became the
 attitude-dynamics analogues of the Delaunay variables emerging in the
 theory of orbits. Later Serret \cite{serret} wrote down the explicit form
 for the generating function responsible for the canonical transformation
 from $(\psi,\theta,\phi,\Psi,\Theta,\Phi)$ to Richelot's constants. His
 treatment was further polished by Radau \cite{radeau} and explained in
 detail by Tisserand \cite{tisserand}

\begin{center}
 \begin{figure}
 \includegraphics[width=6in]{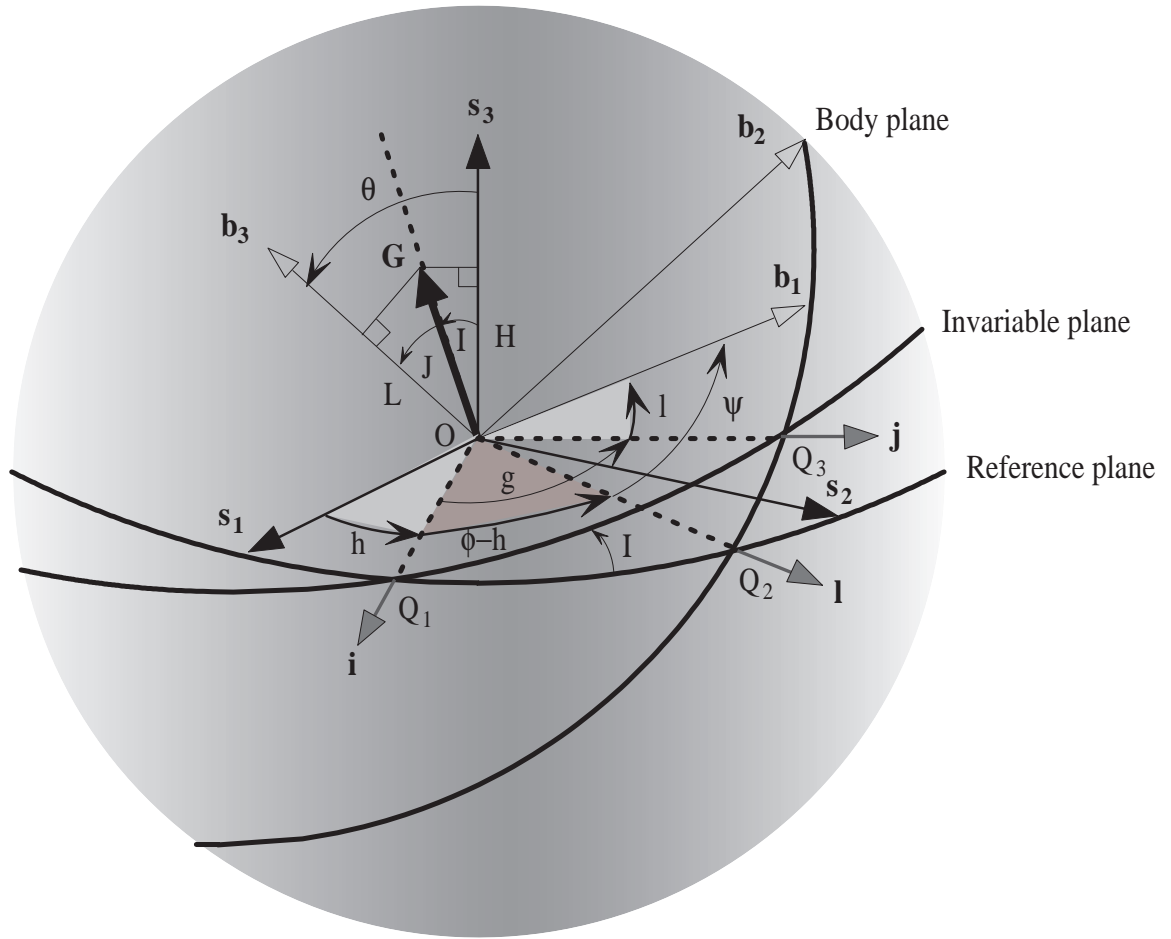}
 \caption{An inertial coordinate system, $\bfs_1,\bfs_2,\bfs_3$, a
 body-fixed frame, $\bfb_1,\bfb_2,\bfb_3$, an angular momentum-based
 frame and intersections of their fundamental planes, denoted by
 $\bfi,\bfl,\bfj$.}
 \label{frames}
 \end{figure}
 \end{center}

\begin{center}
 \begin{figure}
 \includegraphics[width=5.5in]{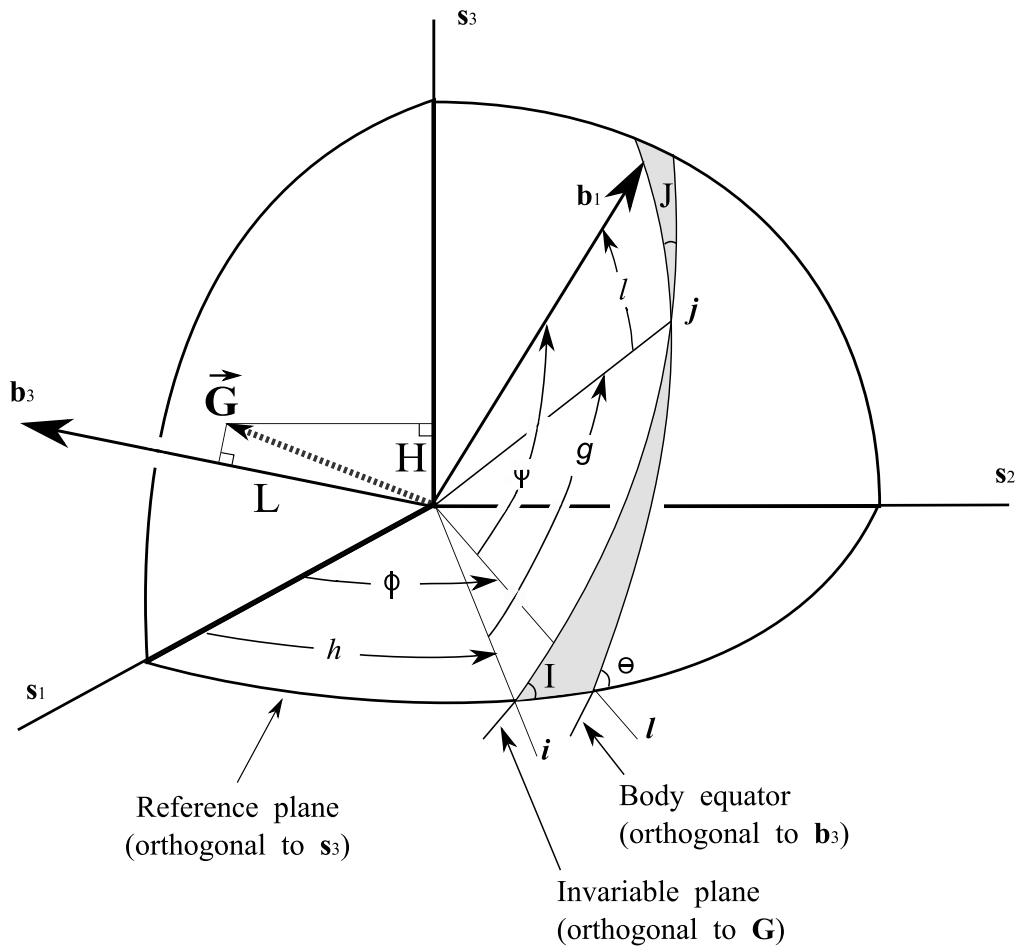}
  \caption{
  The same as the previous picture, but with fewer details.
  The reference coordinate system (inertial or, more generally, precessing)
  is constituted by axes
  $\,{\bf{s}}_1\,,\,{\bf{s}}_2\,,\,{\bf{s}}_3\,$. A body-fixed
  frame is defined by the principal axes
  $\;{\bf{b}}_1\,,\,{\bf{b}}_2\,,\,{\bf{b}}_3\,$.
  The third frame is constituted by the
  angular-momentum vector $\,{{\textbf{G}}}\,$ and a
  plane orthogonal thereto (the so-called invariable plane).
  The lines of nodes are
  denoted with $\,{\bf{i}},\,{\bf{l}},\,{\bf{j}}\,$. The
  attitude of the body relative to the reference frame is
  given by the Euler angles $\,\phi\,,\;\theta\,,\;\psi\,$.
  The orientation of the invariable plane with respect to
  the reference frame is determined by the angles $\,h\,$ and
  $\,I\,$. The inclination $\,I\,$ is equal to the angle that
  the angular-momentum vector $\,{{\textbf{G}}}\,$ makes
  with the reference axis $\,{\bf{s}}_3\,$.
   The angle $\,J\,$ between the invariable plane and the body
  equator coincides with the angle
  that $\,{{\textbf{G}}}\,$ makes with the major-inertia
  axis $\,{\bf{b}}_3\,$ of the body. The projections of the
  angular momentum toward the reference axis $\,{\bf{s}}_3\,$
  and the body axis $\,{\bf{b}}_3\,$ are $\,H\,=\,G\,\cos I\,$
  and $\,L\,=\,G\,\cos J\,$.
  }
 \label{SA_Fig_3.eps}
 \end{figure}
 \end{center}

 The canonical transformation, undertaken by Serret,
 \begin{eqnarray}
 \nonumber
 \left(\;q_1\;\equiv\;\phi\,,\;q_2\;\equiv\;\theta\,,\;q_3\;\equiv\;\psi\,,
 \;p_1\;\equiv\;\Phi\,,\;p_2\;\equiv\;\Theta\,,\;p_3\;\equiv\;\Psi\;;\;\,
 {\cal H}(q,\,p)\;\right)\;
 \rightarrow\\
 \nonumber\\
 \left(\;Q_1\,,\;Q_2\,,\;Q_3\,,\;P_1\,,\;P_2\,,\;P_3\;;\;\,{\cal H}^*
 (Q,\,P)\;\right)\;\;\;,~~~~~~~~~~~~~~~~~~~~~~~~~~~~
 \label{36}
 \label{cantran}
 \end{eqnarray}
 is based on the fact already known to the mathematicians of the second part of the XIX$^{th}$
 century: since both descriptions, $\;(\,\mathbf{q},\,\mathbf{p},\;{\cal H}(\mathbf{q},
 \mathbf{p})\,)\;$ and
 $\;(\,\mathbf{Q},\,\mathbf{P},\,{\cal H}^*(\mathbf{Q},\mathbf{P})\,)\;$, are postulated to
 satisfy the Hamiltonian equations, then the infinitesimally small quantities
 \ba
 d\zeta\;=\;\mathbf{p}^T\;d\mathbf{q}\;-\;{\cal H}\;dt\;\;\;\;\,
 \label{37}
 \ea
and
 \ba
 d{\tilde\zeta}\;=\;\mathbf{Q}^T\;d\mathbf{P}\;+\;{\cal H}^*\;dt\;\;\;,
 \label{38}
 \ea
are perfect differentials, and so is their sum
 \ba
 dS\;\equiv\;d\zeta\;+\;d{\tilde\zeta}=\;\mathbf{p}^T\;d\mathbf{q}\;+\;
 \mathbf{Q}^T\;d\mathbf{P}\;-\;\left(\,{\cal H}\;-\;{\cal H}^*\right)\;dt\;
 \;\;.
 \label{39}
 \ea
 If we start with a system described with $\;(\,\mathbf{q},\,\mathbf{p},\,
 {\cal H}(\mathbf{q},\mathbf{p})\,)\;$, it is worth looking for such a
 re-parameterisation $\;(\,\mathbf{Q},\,\mathbf{P},\,{\cal H}^{*}(\mathbf{
 Q},\mathbf{P})\,)\;$ that the new Hamiltonian $\;H^{*}\;$ is constant in
 time, because this will entail simplification of the canonical equations
 for $\,\mathbf{Q}\,$ and $\,\mathbf{P}\,$. Especially convenient would be to
 find a transformation that would nullify the new Hamiltonian $\;{\cal{H}}^{*}
 \;$, for in this case the new canonical equations would render the
 variables $\;(\,\mathbf{Q},\,\mathbf{P}\,)\;$ constant. One way of seeking
 such transformations is to consider $\;S\;$ as a function of only $\;
 \mathbf{q}\;$, $\;\mathbf{P}\;$, and $\;t\,$. Under this assertion, the
 above equation will result in
 \be
 \frac{\partial S}{\partial t}\;dt\;+\;\frac{\partial S}{\partial
 \mathbf{q}}\;d\mathbf{q}\;+\;\frac{\partial S}{\partial \mathbf{P}}\;d
 \mathbf{P}\;=\;\mathbf{p}^T\,d\mathbf{q}\;+\;\mathbf{Q}^T\,d\mathbf{P}\;-
 \;\left({\cal H}\,-\,{\cal{H}}^{*}\right)\,dt~~~~~
 \label{40}
 \ee
whence
 \be
 \mathbf{p}~=~\frac{\partial S}{\partial \mathbf{q}}~\;\;\;,\;\;\;\;\;\;\;
 \mathbf{Q}~=~\frac{\partial S}{\partial \mathbf{P}}~\;\;\;,\;\;\;\;\;\;\;
 {\cal H}\;+\;\frac{\partial S}{\partial t}\;=\;{\cal H}^{*}\;\;\;.
 \label{41}
 \ee
 The function $\;S(\mathbf{q}\,,\;\mathbf{P}\,,\;t)\;$ can be then found
 by solving the Jacobi equation
 \ba
 {\cal H}\left(\mathbf{q},\,\frac{\partial S}{\partial \mathbf{q}}\,,\,t
 \right)\;+\;\frac{\partial S}{\partial t}\;=\;{\cal H}^{*}\left(\frac{
 \partial S}{\partial \mathbf{P}}\,,\,P\,,\,t\right)\;\;\;.
 \label{42}
 \label{jacobiequation}
 \ea
 In the free-spin case, the Jacobi equation becomes
 \ba
 \nonumber
 \frac{1}{2 \sin^2{\theta }}\left(\frac{\sin^2{\psi }}{I_1}+\frac{\cos^2{\psi }}{I_2}\right)
 \left({\frac{\partial\mathcal S}{\partial\phi}-
 \frac{\partial\mathcal S}{\partial\psi} \cos{\theta}}\right)^2+
 \frac{1}{2}\left(\frac{\cos^2{\psi }}{I_1}+\frac{\sin^2{\psi }}{I_2}\right)
 \left(\frac{\partial\mathcal S}{\partial\theta}\right)^2\;+
 \ea
 ~\\
 \ba
 \frac{1}{2I_3}
 \left(\frac{\partial\mathcal S}{\partial\psi}\right)^2+\left(\frac{1}{I_1}-\frac{1}{I_2}
 \right)\frac{\partial\mathcal S}{\partial\theta}\left(
{\frac{\partial\mathcal S}{\partial\phi}-\frac{\partial\mathcal
S}{\partial\psi}\cos{\theta}}\right)\frac{\sin{\psi}\,\cos{\psi}}{\sin{\theta}}
+\frac{\partial\mathcal S}{\partial t}={\cal
H}^*\left(\frac{\partial S}{\partial \mathbf{P}}\,,
 \,\mathbf{P}\,,\,t\right) ~~.~~
 \label{43}
 \label{hamil}
 \ea
 At this point, Serret \cite{serret}, Radau \cite{radeau}, and Tisserand \cite{tisserand}
 chose to put $~{\cal H}^*~$ equal to zero, thereby predetermining the new variables $~(\,
 \mathbf{Q}\,,~\mathbf{P}\,)~$ to come out constants. (See, for example, equation (21) on
 page 382 in \cite{tisserand}.) Thence the Jacobi equation (\ref{jacobiequation}) became
 equivalent to\footnote{~The generating function should be written down exactly as (\ref{44})
 and \textbf{not} as
  \ba
  \nonumber
 {\cal H}\left(\mathbf{q},\,\frac{\partial S}{\partial \mathbf{q}}\,\right)\;+\;
 \frac{d S}{d t}\;-\;\frac{\partial S}{\partial q}\;\frac{d q}{dt}\;-\;\frac{\partial S}{
 \partial P}\;\frac{d P}{dt}\;=\;0\;\;\;,
 \label{}
 \ea
 because the new momenta $\;P\;$ are not playing the role of
 independent variables but will emerge as constants
 in the solution.}
 \be
 {\cal H}\left(\mathbf{q},\,\frac{\partial S}{\partial \mathbf{q}}\,\right)\;+\;
 \frac{d S}{d t}\;-\;\frac{\partial S}{\partial q}\;\frac{d q}{dt}\;=\;0\;\;\;.
 \label{44}
 \ee
 By taking into account that the initial Hamiltonian $\,\cal H\,$ depends explicitly
 neither on the time nor on the angle $\,\phi\,$, Serret and his successors granted themselves
 an opportunity to seek the generating function in the simplified form of
 \ba
 S\;=\;A_1\;t\;+\;A_2\;\phi\;+\;\int \frac{\partial S}{\partial \theta}\; d\theta \;+\;\int
 \frac{\partial S}{\partial \psi}\;d\psi\;+\;C\;\;\;,
 \label{45}
 \ea
the constant $\;A_1\;$ being equal to the negative value of the
Hamiltonian $\;{\cal H}\;$, i.e., to the negative rotational
kinetic energy:
 \ba
 A_1\;=\;-\;T_{kin}\;\;\;.
 \label{46}
 \ea
 The second constant, $~A_2~$, as well as the derivatives $~{\partial S}/{\partial \theta}~$
 and $\;{\partial S}/{\partial\psi}\;$ can be calculated via the first formula (\ref{41}):
 \ba
 A_2\;\equiv\;\frac{\partial S}{\partial
 \phi}\;=\;\Phi
 \;\;\;,\;\;\;\;
 \frac{\partial S}{\partial \theta}\;=\;\Theta
 \;\;\;,\;\;\;\;
 \frac{\partial S}{\partial \psi}\;=\;\Psi
 \;\;\;,\;\;\;
 \label{47}
 \ea
  so we get:
 \ba
 S\;=\;-\;t\;T_{kin}\;+\;\Phi\;\phi\;+\;\int \Theta\; d\theta \;+\;\int
 \Psi\;d\psi\;+\;C\;\;\;,
 \label{48}
 \ea
 $\Phi\,$, $\,\Theta\,$, and $\,\Psi\,$ being given by formulae (\ref{33} - \ref{35}),
 and $\,\Phi\,$ being a constant of motion because $\,{\cal H}\,$ is
 $\,\phi$-independent. Plugging of (\ref{33} - \ref{35}) into
 (\ref{48}) yields:
 \ba
 \nonumber
 S\,=\;-\;t\;T_{kin}\,+\;\Phi\;\phi\;+\;
 G\;\int \cos{J}\;d\psi\;+\;G\;\int
 \sin{J}\,\;\sin{\left(\,l\,-\;\psi
 \,\right)}\;d\theta\;+\;C
 \ea
 \ba
 \nonumber\\
 \nonumber
 =\;-\;t\;T_{kin}\;+\;G\;\phi\;\cos I\;+\;
 G\;\int \cos{J}\,\;dl\,+~~~~~~~~~~~~~~~~~~~~~~~~~~~\\
 \nonumber\\
 G\;\int \left[\,-\;\cos{J}\,\;d\left(\,l\,-\;\psi
 \,\right)\;+\;\sin{J}\,\;\sin{\left(\,l\,-\;\psi \,\right)}\;d\theta
 \,\right]\;+\;C
 \;\;,~~
 \label{49}
 \ea
 The latter expression can be simplified through the equality (derived in the Appendix A.1)
 \ba
 d \;g \;-\;d\;(\phi-h)\;\,\cos I\;=\;-\;\cos{J}\;\,d\left(l\,-\;\psi
 \right)\;+\;\sin{J}\;\,\sin{(l - \psi)}\;d\theta\;\;\;
 \label{50}
 \ea
 the angles $\;(\phi-h)\;$ and $\;g\;$ being shown on Fig.~\ref{frames}, and
 Fig.~\ref{SA_Fig_3.eps}. This
 will entail:
 \ba
 S\;=\;-\;t\;T_{kin}\;+\;G\;h\;\cos I\;+\;
 G\;g\;+\;G\;\int \cos{J}\,\;dl\,+\;C\;\;,\;\;\;
 \label{51}
 \ea
 where the constant $\,C\,$ may be chosen, for example, as zero.
 At this point, the authors of the XIX$^{th}$ century made use of two
 observations. First, in the unperturbed case, one can introduce
 the dimensionless time
 \ba
 u\;=\;n\;(t\,-\,t_o)\;\;\;,
 \label{dimensionless}
 \ea
 $n\,\equiv\,\sqrt{\left(\,G^2\,I_1^{-1}\,-\,2\,T_{kin}\,\right)\,
 \left(\,I_2^{-1}\,-\,I_3^{-1}\,\right)}\;$ being the mean angular velocity
 (or, playing an astronomical metaphor, the ``mean motion"). At each instance
 of time, the value of so-defined $\,u\,$ is unambiguously determined by the
 instantaneous attitude of the rotator, via the solution of the equations of
 motion (this solution is expressed through the elliptic functions of {\it{\,u}}).
 More specifically, $\,u\,$ is a function of $\;T_{kin}\,,\;\,G\,,\;\,G\cos I\,,
 \;\,\theta\,,\;\mbox{and}\;\psi\;$ -- the appropriate expression is given in
 Appendix A.2 below. Simply from this definition it trivially follows that in
 the unperturbed case the difference $\;u/n\,-\,t\;$ is an integral of motion,
 because it is identically nil. Under perturbation, though, the
 afore mentioned function $\;u(T_{kin}\,,\;G\,,\;G\cos I\,,\;\theta\,,\;\psi)\;$,
 divided by $\,n\,$, will no longer furnish the actual values of the physical
 time $\;t\;$. (It would, had we substituted in (\ref{dimensionless}) the product
 with an integral.) Hence, under perturbation, the integral of
 motion $\;u/n\,-\,t\;$ will acquire time dependence.

 The second observation can be done if we
 consider a coordinate system defined by the angular-momentum vector
 and a plane perpendicular thereto. This plane (in astronomy, called
 {\it{invariable}}) will be chosen as in Fig.~\ref{frames} and
 Fig.~\ref{SA_Fig_3.eps}, so that the centre of mass of the body lie
 in this plane. The Euler angles defining the attitude of a body relative
 to the invariable plane are $\;g\,,\;J\,,\;l\;$, as on Fig.~\ref{frames},
 and Fig.~\ref{SA_Fig_3.eps}. In the case of free spin, they obey the
 differential relation:\footnote{~Tisserand (\cite{tisserand}, p.386)
 and his contemporaries used to prove (\ref{52}) by a somewhat tedious method
 based on formulae connecting the angles $\;J\,,\;l\;$ with the components
 of the body angular velocity $\;\omegabold\;$ and with the absolute value of the
 angular momentum:
 \ba
 \nonumber
 I_1\,\omega_1\;=\;G\;\sin J \; \sin l  \;\;\;,\;\;\;\;\;
 I_2\,\omega_2\;=\;G\;\sin J \; \cos l  \;\;\;,\;\;\;\;\;
 I_3\,\omega_3\;=\;G\;\cos J \;\;\;.\;\;\;\;\;
 \ea
 We provide this proof in the Appendix A.3 below.

 The Andoyer variables, which we shall introduce below, make (\ref{52}) self-evident:
 (\ref{52}) follows from equations (\ref{cbcp}), (\ref{dot_g}), (\ref{dot_l}) and
 (\ref{newhamil}).}
  \ba
  d g\;+\;\cos{J}\;dl\;=\;\frac{2\;T_{kin}}{G}
  \;dt\;\;\;
  \label{52}
  \ea
  whose integration results in
 \ba
 g\;-\;\frac{2\;T_{kin}}{G}\;\frac{u}{n}\;+\;\int\,\cos
 J\,dl\,=\;g_o\;\;\;.
 \label{54}
 \ea
 Trivially, the expression standing in the left-hand side is an integral of
 motion in the unperturbed case,  because in this case it is equal to
 $\;g_o\;$. However, under disturbance the left-hand side will no longer be
 equal to $\;g_o\;$ (and thus will no longer remain an integral of motion),
 because neither (\ref{dimensionless}) nor (\ref{52}) will hold.

  The above two observations motivate the following equivalent
  transformation of (\ref{51}):
  \ba
  S~=~\left(\,2~\frac{u}{n}~-~t\,\right)~T_{kin}~+~Gh~\cos
  I\;\;+\;G\;g_o\;\;.\;\;\;\;
  \label{55}
  \ea
 Now it might be tempting to state that, if we choose
 $~(\,P_1\,,~P_2\,,~P_3\,)\,\equiv\,(\,G\,,~
 G\,\cos I\,,\;T_{kin}\,)\;$, then expression (\ref{55}) would entail
 \ba
 \frac{\partial S}{\partial G}\;=\;Q_1
 \label{56}
 \ea
and
 \ba
 \frac{\partial S}{\partial \left(G\,\cos I\right) }\;=\;Q_2
 \label{57}
 \ea
 where
 \ba
 Q_1\;\equiv\;g_o
 \label{58}
 \ea
and
 \ba
 Q_2\;\equiv\;h
 \label{59}
 \ea
are integrals of motion. These two formulae, if correct, would
implement our plan set out via the second equation (\ref{41}).

In fact, equations (\ref{55} - \ref{56}) do {\textbf{not}}
immediately follow from (\ref{54}), because the variations of
$\;G\,,\;G\,\cos I\,,\;T_{kin}\;$ are \textbf{not} independent from
the variations of the angles involved. These variations are subject
to some constraints. A careful calculation, presented in the
Appendix below, should begin with writing down an expression for
$\;\delta S\;$, with the said constraints taken into account. This
will give us:
 \ba
 \delta S\;=\;\left(\,\frac{u}{n}\;-\;t\,\right)\;\delta T_{kin}\;+\;
 h\;\delta (\,G\;\cos I\,)\;+\;g_o\;\delta G\;\;\;
 \label{60}
 \ea
whence we deduce that (\ref{56} - \ref{59}) indeed are correct. We
also see that the negative of the initial instant of time (which,
too, is a trivial integral of motion), turns out to play the role of
the third new coordinate:
  \ba
 \frac{\partial S}{\partial T_{kin}}\;=\;Q_3
 \label{61}
 \ea
where
 \ba
 Q_3\;\equiv\;\frac{u}{n}\;-\;t\;=\;-\;t_o\;\;\;.
 \label{62}
 \ea
All in all, we have carried out a canonical transformation from the
Euler angles and their conjugate momenta to the Serret variables
$\;Q_1\,=\,g_o\;$, $\;Q_2\,=\,h \;,
 \;Q_3\,=\,-\;t_o\,$ and their conjugate momenta $\,P_1\,=\,G\,$, $P_2\,=\;G\;\cos I
\,,\;P_3\,=\;T_{kin}\;$. In the modern literature, the constant
$\;P_2\;$ is denoted by $\;H\;$. Constants $\;h\,,\;G\,,\;H\;$
enter the Andoyer set of variables, discussed in the next
subsection.

 \subsection{Andoyer (1923), and Deprit \& Elipe (1993)}


While Serret \cite{serret} had come up with a full reduction of the problem, Andoyer
\cite{andoyer} suggested a partial reduction, based on a general method of obtaining a
set of canonical variables ($n$ coordinates and their $n$ conjugate moments) from a set
of $2n$ variables that are not necessarily canonical. This general method is discussed
in the Appendix, \S A.5.

To understand the essence of that partial reduction, we shall start
with a transformation from the inertial to the body frame via a
coordinate system associated with the invariable plane.

Referring to Fig.~\ref{frames}, let $OQ_1$ and $OQ_3$ denote the
LON's obtained from the intersection of the invariable plane with
the inertial plane and with a plane fixed within the body,
respectively. Let $\bfi$ be a unit vector along the direction of
$OQ_1$, and $\bfj$ be a unit vector along $OQ_3$. Define a
$3-1-3-1-3$ rotation sequence as follows:
\begin{itemize}
\item $R(h,\bfs_3)$, a rotation about $\bfs_3$ by $0\leq h < 2\pi$,
mapping $\bfs_1$ onto $\bfi$.

\item $R(I,\bfi)$, a rotation about $\bfi$ by $0 \leq I
<\pi$, mapping $\bfs_3$ onto the angular momentum vector, $\bfG$.

\item $R(g,\bfG/G)$, a rotation about a unit vector pointing in the
direction of the angular momentum by $0 \leq g < 2\pi$, mapping
$\bfi$ onto $\bfj$.

\item $R(J,\bfj)$, a rotation about $\bfj$ by $0 < J < \pi$,
mapping $\bfG$ onto $\bfb_3$.

\item $R(\l,\bfb_3)$, a rotation about $\bfb_3$ by $0 \leq l \leq
2\pi$, mapping $\bfj$ onto $\bfb_1$.
\end{itemize}
The composite rotation may be written as
 \begin{equation}
 \label{Rnew}
    R(h,I,g,J,l)=R(\l,\bfb_3)R(J,\bfj)R(g,\bfG/G)R(I,\bfi)R(h,\bfs_3)
 \end{equation}
Evaluation of the product of these five matrices will result in
 \begin{equation}
 \label{Rnewmat}
 R(h,I,g,J,l) = \left[\bfv_1,\;\bfv_2,\;\bfv_3\right]
 \end{equation}
where
 \begin{equation}\label{v1}
    \bfv_1 = \left[%
 \begin{array}{c}
  (\cc_l\cc_g-\s_l\cc_J\s_g)\cc_h-(\cc_l\s_g+\s_l\cc_J\cc_g)\cc_I-\s_l\s_J\s_I\s_h \\
  -\s_l\cc_g-\cc_l\cc_J\s_g\cc_h-(-\s_l\s_g+\cc_l\cc_J\cc_g)\cc_I-\cc_l\s_J\s_I\s_h \\
  \s_J\s_g\cc_h+(\s_J\cc_g\cc_I+\cc_J\s_I)\s_h \\
 \end{array}%
 \right]
 \end{equation}
 \begin{equation}
 \label{v2}
    \bfv_2 = \left[%
 \begin{array}{c}
  (\cc_l\cc_g-\s_l\cc_J\s_g)\s_h+(\cc_l\s_g+\s_l\cc_J\cc_g)\cc_I-\s_l\s_J\s_I\cc_h \\
  -\s_l\cc_g-\cc_l\cc_J\s_g\s_h+(-\s_l\s_g+\cc_l\cc_J\cc_g)\cc_I-\cc_l\s_J\s_I\cc_h \\
  \s_J\s_g\s_h+(-\s_J\cc_g\cc_I-\cc_J\s_I)\cc_h \\
 \end{array}%
 \right]
 \end{equation}
 \begin{equation}\label{v3}
    \bfv_3 = \left[%
 \begin{array}{c}
  (\cc_l\s_g+\s_l\cc_J\cc_g)\s_I+\s_l\s_J\cc_I \\
  -\s_l\s_g+\cc_l\cc_J\cc_g\s_I+\cc_l\s_J\cc_I\\
  -\s_J\cc_g\s_I+\cc_J\cc_I\\
 \end{array}%
 \right]
 \end{equation}

A sufficient condition for the transformation (\ref{cantran}) to be
canonical can be formulated in terms of perfect differentials. Given
that the Hamiltonian lacks explicit time dependence, this condition
will read\footnote{We shall show shortly that indeed the perfect
differentials criterion for canonical transformations holds in this
case with the perfect differential of the generating function being
identically equal to zero.} \cite{schaubjunkins}:
 \ba
 \label{cantran2}
    \Phi d\phi+\Theta d\theta + \Psi d\psi = Ldl+Gdg+Hdh
 \ea
Let us first evaluate the left-hand side of (\ref{cantran2}). The
differential of $R(\phi,\theta,\psi)$ is readily found to be
\cite{leubner}
 \ba
 \label{dr1}
 dR(\phi,\theta,\psi) = \bfs_3d\phi+\bfl d\theta+\bfb_3d\psi
 \ea
Multiplying both sides of (\ref{dr1}) by $\bfG$, and taking
advantage of the identities (cf.~(\ref{G1} - \ref{G3})), one will
arrive at
 \ba
    \bfG\;\cdot\;\bfs_3\;=\;\Phi\;\;,\;\;\;\;
    \bfG\;\cdot\;\bfl\;=\;\bfG\;\cdot\;(\bfs_1\cc_\phi+\bfs_2\s_\phi)=\Theta\;\;,\;\;\;\;
    \bfG\;\cdot\;\bfb_3\;=\;\Psi\;\;,\;\;\;
    \label{}
 \ea
whence
 \ba
 \nonumber
 \Phi \; d\phi\;+\;\Theta \;d\theta \;+\; \Psi d\psi
 \;=~~~~~~~~~~~~~~~~~~~~~~~~~~~~~~~~~~~~~~~~~~~~~~~~~~~~~
 \ea
 \ba
 \bfG\;\cdot\;\bfs_3\;d\phi\;+\;\bfG\;\cdot\,(\bfs_1\,\cc_\phi\,+\,\bfs_2\,\s_\phi)\;d\theta\;+\;
 \bfG\;\cdot\;\bfb_3\;d\psi\;=\;{\bf G}\;\cdot\;dR\;\;\;.
 \label{471}
 \ea
We shall now repeat the above procedure for the right-hand side of
(\ref{cantran2}). The differential of $R(h,I,g,J,l)$ is evaluated
similarly to (\ref{dr1}) so as to get
 \ba
 \label{dr2}
 dR(h\,,\;I\,,\;g\,,\;J\,,\;l)\;=\;\bfs_3 \;dh\;+\;\bfi\;
 dI\;+\;\frac{\bfG}{G}\; dg\;+\;\bfj \;dJ\;+\; \bfb_3\; dl
 \ea
 Since
 \ba
 \label{}
    \bfG\cdot\bfs_3=\Phi,\quad
    \bfG\cdot\bfi=0,\quad\bfG\cdot\bfG/G=G,\quad
    \bfG\cdot\bfj=0,\quad \bfG\cdot\bfb_3=\Psi\;\;\;,
 \ea
 then multiplication of both sides of (\ref{dr2}) by $\bfG$ will lead us to
 \begin{eqnarray}
    \bfG\cdot dR = \Phi dh+ G dg+\Psi dl
     \label{474}
     \label{74}
 \end{eqnarray}
Together, (\ref{471}) and (\ref{474}) give
 \ba
 \Phi\;d\phi\;+\;\Theta\;d\theta\;+\;\Psi\;d\psi\;=\;{\bf G}\,\cdot\,dR\;= \;\Phi\;dh\;+\;G\;
 dg\;+\;\Psi\; dl\;\;\;,
 \label{75}
 \ea
comparison whereof with (\ref{cantran2}) immediately yields
 \begin{eqnarray}
 \label{eqmom}
    \Phi=H\;\;\;,\quad \Psi = L\;\;\;.
    \label{76}
 \end{eqnarray}
To conclude, the transition from the Euler coordinates
$\;\phi\,,\;\theta\,,\;\psi\;$ to the coordinates
$\;h\,,\;g\,,\;l\;$ becomes a canonical transformation (with a
vanishing generating function, as promised above), if we choose
the momenta $\;H\,,\;G\,,\;L\;$ conjugated to $\;h\,,\;g\,,\;l\;$
as $\;H\,=\,\Phi\,,\;G\,=\,|{\bf G}|\,,\;L\,=\,\Psi\;$,
correspondingly.

To get direct relations between the Euler angles and Andoyer angles
$\;h\,,\;g\,,\;l\;$, we can compare the entries of
$R(\phi,\theta,\psi)$ and $R(h,I,g,J,l)$. Alternatively, we may
utilise the spherical laws of sines and cosines written for the
spherical triangle $Q_1 Q_2 Q_3$ \cite{kinoshita77}. Adopting the
former approach, we first express $I$ and $J$ in terms of the SA
canonical momenta (cf. Fig.~\ref{frames}),
 \ba
    \cc_I=H/G\;\;,\;\;\;\quad \cc_J =L/G\;\;.
 \label{cbcp}
 \label{84}
 \ea
Equating the $(3,3)$ entries, the quotient of the $(1,3)$ and
$(2,3)$ entries, and the quotient of the $(3,1)$ and the $(3,2)$
entries yields, respectively,
 \ba
 \label{ct}
 \cc_\theta = \frac{LH}{G^2}-\frac{\cc_g}{G^2}\sqrt{(G^2-L^2)(G^2-H^2)}
 \label{78}
 \ea
 \ba
 \label{tp}
 \tan \psi = \frac{\sqrt{G^2-H^2}(G\s_g\cc_l+L\s_l\cc_g)+
 H\s_l\sqrt{G^2-L^2}}{-\sqrt{G^2-H^2}(G\s_g\s_l+L\cc_l\cc_g)+H\cc_l\sqrt{G^2-L^2}}
 \label{79}
 \ea
 \ba
 \label{th}
 \tan \phi =
 -\frac{\sqrt{G^2-L^2}(G\s_g\cc_h+H\s_h\cc_g)+L\s_h\sqrt{G^2-H^2}}{\sqrt{G^2-L^2}(G\s_g\s_h-
 H\cc_h\cc_g)-L\cc_h\sqrt{G^2-H^2}}
 \label{80}
 \ea

To complete the equations of transformation relating the Eulerian
variables to the Andoyer variables, we utilise either
Eqs.~(\ref{g1})-(\ref{g3}) or Eqs.~(\ref{G1})-(\ref{G3}) to
calculate the magnitude of $\bfG$ and then express the momentum
$\Theta$ in terms of the Andoyer variables. Carrying out this
procedure yields
 \ba
 \label{Theta}
 \Theta = G\s_J\s_{\psi-l}=\sqrt{G^2-L^2}\s_{\psi-l}
 \ea
where the expressions for $\cc_\psi$ and $\s_\psi$ may be readily
obtained from Eq. (\ref{tp}).

 Substituting the identities (\ref{eqmom}) and (\ref{ct} - \ref{Theta}) into
 (\ref{23}) yields the new, single-degree-of-freedom Hamiltonian
 \ba
 \label{newhamil}
 \mathcal H(g,h,l,G,H,L)=\frac{1}{2}\left(\frac{\s_l^2}{I_1}+\frac{\cc_l^2}{I_2}\right)(G^2-
 L^2)+\frac{L^2}{2I_3}
 \ea
Alternatively, we could have utilised relations (\ref{32a}) and (\ref{84}), to
get
  \ba
  g_1 &=& I_1\omega_1 \;=\; G\s_J\s_l = \sqrt{G^2-L^2}\s_l \label{g1}
  \label{1490}\;\;\;,\\
  g_2 &=& I_2\omega_2 \;=\; G\s_J\cc_l = \sqrt{G^2-L^2}\cc_l \label{g2}\;\;\;,\\
  \label{1491}
  g_3 &=& I_3\omega_3 \;=\; L \label{g3}\;\;\;,
  \label{1492}
  \ea
 insertion whereof into (\ref{kinetic}) would then lead us to the same result
 (\ref{newhamil}).

We would note that in the new Hamiltonian the coordinates $g,h$ are cyclic,
and hence the momenta $G,H$ are integrals of motion. Also, since
\begin{equation}\label{constraint}
-G\le L \le G,
\end{equation}
the hamiltonian is non-negative,
\begin{equation}\label{hge0}
    \mathcal H \geq 0
\end{equation}

\begin{center}
\begin{figure}
\includegraphics[width=6in]{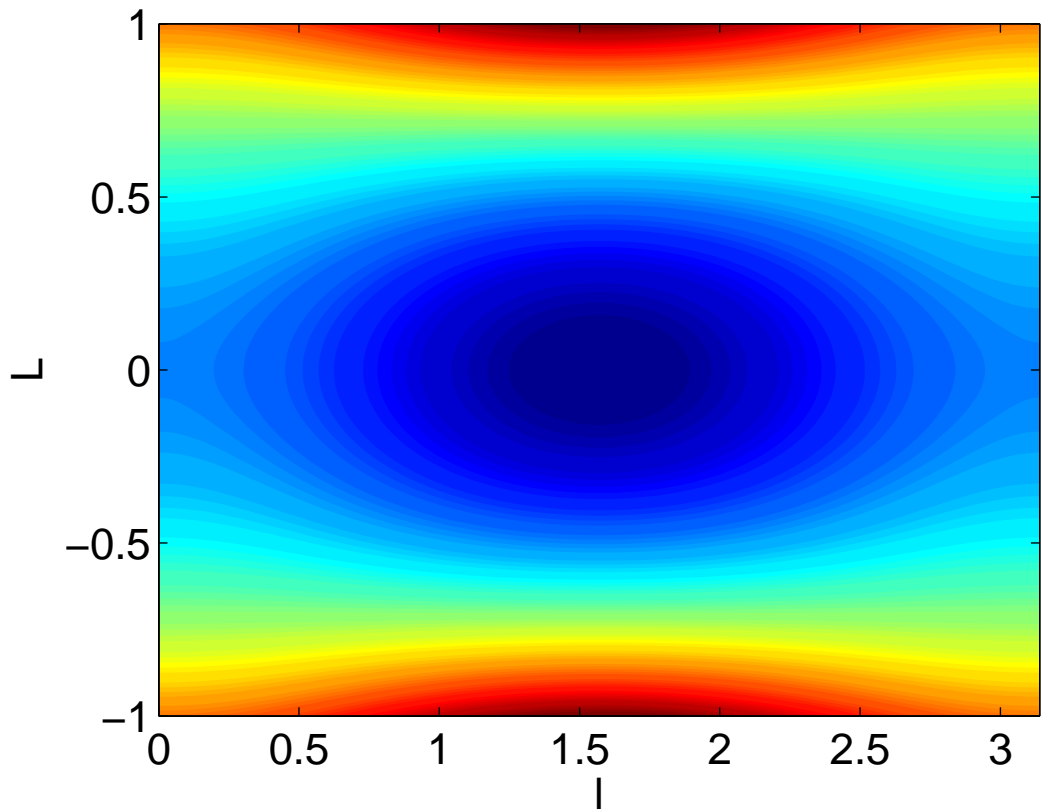}
\caption{The phase plane of $(l,L)$ comprising the isoenergetic
curves of the Serret-Andoyer free-motion Hamiltonian.}
\label{fig:hamil}
\end{figure}
\end{center}

To get the canonical equations of motion, denote the generalised coordinates
by ${\mbox{\bf q}}=[g,h,l]^T$ and the conjugate momenta by $ \mbox{\bf p} =
[G,H,L]^T$.
Hamilton's equations in the absence of external torques are
\begin{eqnarray}
  \dot{\bf q} &=& \frac{\partial\mathcal H}{\partial {\bf p}} \\
  \dot{\bf p} &=& -\frac{\partial\mathcal H}{\partial {\bf q}}
\end{eqnarray}
Evaluating Hamilton's equations for (\ref{newhamil}), yields the
canonical equations of free rotational motion:
\begin{eqnarray}
  \dot g &=& \frac{\partial\mathcal H}{\partial G} = G\left(\frac{\sin^2{l}}{I_1}+
  \frac{\cos^2{l}}{I_2}\right)
  \label{dot_g}\\
  \dot h &=& \frac{\partial\mathcal H}{\partial H} = 0
  \label{dot_h}\\
  \dot l &=& \frac{\partial\mathcal H}{\partial L} = L\left(\frac{1}{I_3}-
  \frac{\sin^2{l}}{I_1}-\frac{\cos^2{l}}{I_2}\right) \label{l}
  \label{dot_l}\\
  \dot G &=& -\frac{\partial\mathcal H}{\partial g} = 0
  \label{dot_G}\\
  \dot H &=& - \frac{\partial\mathcal H}{\partial h} = 0
  \label{dot_H}\\
  \dot L &=& - \frac{\partial\mathcal H}{\partial l} =
  (L^2-G^2)\sin{l}\;\cos{l}\left(\frac{1}{I_1}-\frac{1}{I_2}\right)
  \label{dot_L}
  \label{L}
\end{eqnarray}
Eqs.~(\ref{l}), (\ref{dot_L}) are separable differential
equations, and hence can be solved in a \emph{closed form}
utilising the fact that $\mathcal H$ is constant (implying 
conservation of energy) \cite{depritelipe}. The $(l,L)$ phase
plane may be characterised by plotting isoenergetic curves of the
Hamiltonian (\ref{newhamil}). An example plot is depicted by
Fig.~\ref{fig:hamil}, clearly showing the separatrix between
rotational and librational motions.

 \section{THE ~CANONICAL ~PERTURBATION\\ THEORY
          ~IN ~APPLICATION ~TO\\ ~ATTITUDE ~DYNAMICS
          ~AND ~TO\\ ~ROTATION ~OF ~CELESTIAL BODIES}

The content of this section is based mainly on the results of papers
by Efroimsky \cite{efroimsky_2004} and Efroimsky \& Escapa
\cite{efroimsky_2007}, to which we refer the reader for details.

 \subsection{A modified Andoyer set of variables}

\begin{center}
\begin{figure}
\includegraphics[width=5in]{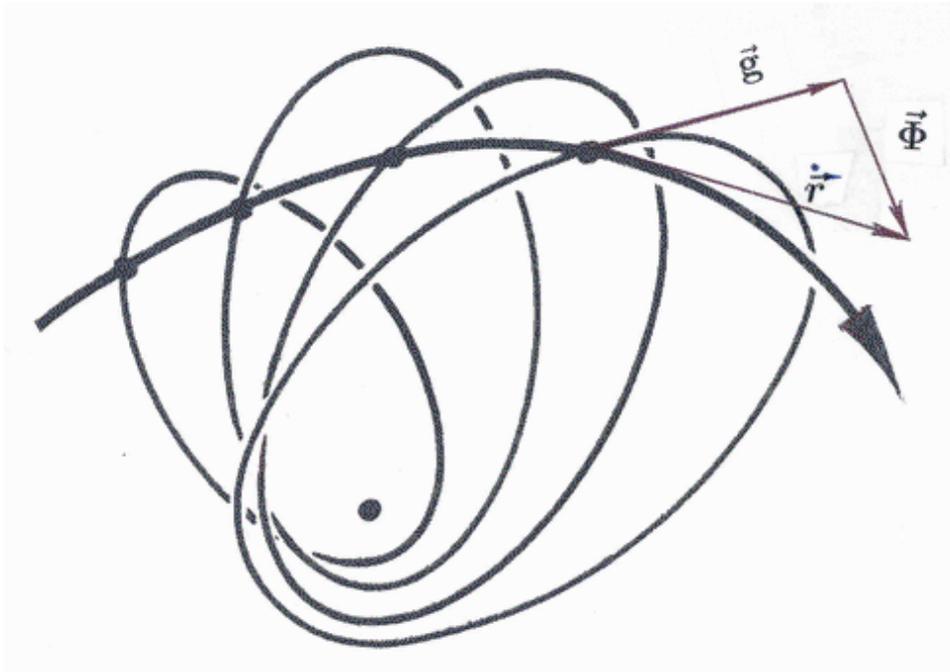}
\caption{The perturbed trajectory is a set of points belonging to a
sequence of confocal instantaneous ellipses. The ellipses are
\textbf{not} supposed to be tangent, nor even coplanar to the orbit
at the intersection point. As a result, the physical velocity
$\,\doterbold\,$ (tangent to the trajectory at each of its points)
differs from the Keplerian velocity $\,\bf g\,$ (tangent to the
ellipse).} \label{fig:michael1}
\end{figure}
\end{center}

\begin{center}
\begin{figure}
\includegraphics[width=5in]{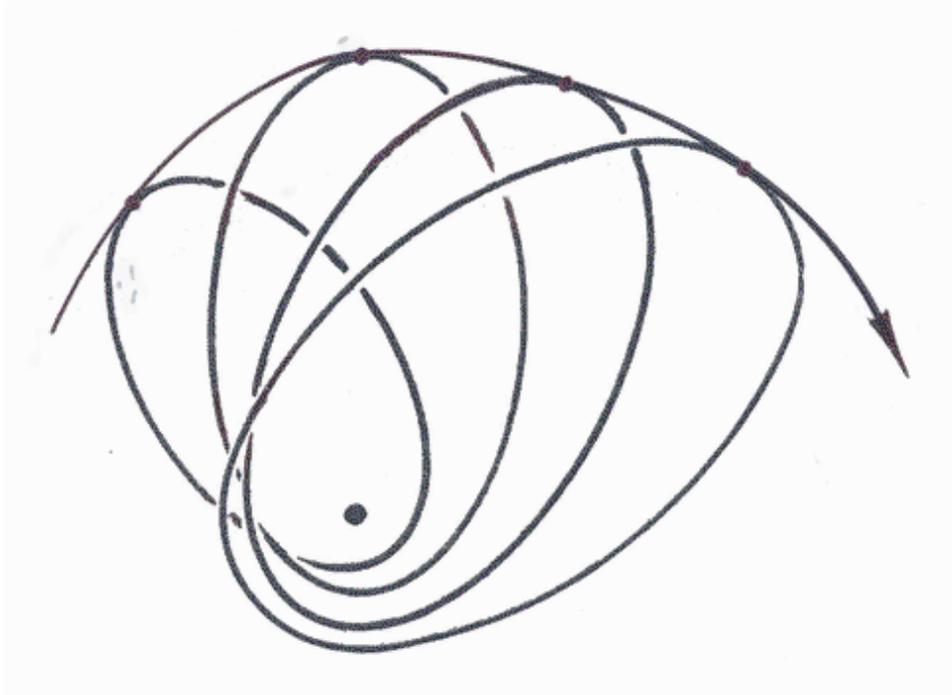}
\caption{The perturbed trajectory is represented through a sequence
of confocal instantaneous ellipses which {are} tangent to the
trajectory at the intersection points, i.e., are osculating. Now,
the physical velocity $\,\doterbold\,$ (which is tangent to the
trajectory) will coincide with the Keplerian velocity $\,\bf g\,$
(which is tangent to the ellipse)} \label{fig:michael2}
\end{figure}
\end{center}

To understand how the SA formalism may be used to model disturbing
torques, let us start with an orbital dynamics analogy. In the
theory of orbits, a Keplerian ellipse or hyperbola, emerging as an
unperturbed two-body orbit, is considered as a sort of ``elementary
motion," so that all the other available trajectories are considered
to be distortions of such conics, distortions implemented via
endowing the orbital parameters $\;C_j\;$ with their own time
dependence. Points of the trajectory can be donated by the
``elementary curves" either in a non-osculating manner, as in
Fig.~\ref{fig:michael1},\footnote{~Historically, the first attempt
of using nonosculating elements dates back to Poincare (1897)
\cite{poincare}, though he never explored them from the viewpoint of
a non-Lagrange constraint choice. (See also Abdullah and Albouy
(2001) \cite{abdullah}, p. 430.) Parameterisation of nonosculation
through a non-Lagrange constraint was offered in Efroimsky (2002a,b)
\cite{e1, e2}.} or in the osculating one, as in
Fig.~\ref{fig:michael2}.

Similarly, in attitude dynamics, a complex spin can be presented as
a sequence of configurations constituted by some ``elementary
rotations." The easiest possibility is to use in this role the
Eulerian cones, i.e., the loci of the spin axis, corresponding to
undisturbed spin states. These are the simple motions exhibited by
an undeformable unsupported rotator with no torques acting on
it.\footnote{~Here one opportunity is to use, as ``elementary"
motions, the non-circular Eulerian cones described by the actual
triaxial top, when this top is unforced. Another opportunity is to
use for this purpose the circular Eulerian cones described by a
dynamically symmetrical top (and to consider its triaxiality as
another perturbation). The results of our further study will be
independent from the choice between these two options.} Then, to
implement a perturbed mode, we shall have to go from one Eulerian
cone to another, just as in Figs.~\ref{fig:michael1} and
\ref{fig:michael2} we go from one Keplerian conic to another.
Accordingly, a smooth ``walk" over the instantaneous Eulerian cones
may be osculating or non-osculating.

The physical torques, the triaxiality of the rotator, and the
fictitious torques caused by the frame noninertiality are among
the possible disturbances causing this ``walk." The latter two
types of disturbances depend not only on the orientation but also
on the angular velocity of the body.

In the theory of orbits, we express the Lagrangian of the reduced
two-body problem in terms of the spherical coordinates
$\;q_j\;=\;\{\,r\,,\;\varphi\,,\;\theta\,\}\;$, then calculate the
momenta $\;p_j\;$ and the Hamiltonian $\;{\cal H}(q,\,p)\;$, and
apply the Hamilton-Jacobi method \cite{plummer} in order to get the
Delaunay constants\footnote{~Be mindful that the more customary
Delaunay variables include $\,M\,$ instead of $\,M_o\,$, and the
three latter elements have opposite signs (and play the role of
coordinates, not momenta). For such variables, the Delaunay
Hamiltonian is the {\emph{negative}} of the actual Hamiltonian
perturbation and is, above all, amended with $\,\mu^2/(2L)\,$.
However, the Hamilton-Jacobi procedure performed with the actual,
physical Hamiltonian perturbation leads exactly to the Delaunay
constants (\ref{96}) -- see section 136 in Plummer (1918).}
 \ba
 \nonumber
\{\,Q_1\,,\;Q_2\,,\;Q_3\;;\;P_1\,,\;P_2\,,\;P_3\,\}\,\equiv\,
~~~~~~~~~~~~~~~~~~~~~~~~~~~~~~~~~~~~~~~~~~~~~~~~~~~~~~~~~~~~~\\
\label{96}\\
 \nonumber
\{\;\sqrt{\mu a}\;\;,\;\;\sqrt{\mu a \left(1\,-\,e^2
\right)}\;\;,\;\;\sqrt{\mu a \left(1\,-\,e^2 \right)}\;\cos
\inc\;\;;\;\;\;-\,M_o\;\;,\;\;-\,\omega\;\;,\;\;-\,\Omega\;\,\} \;,~~~~~\,
\ea
 $\mu\;$ being the reduced mass.

Very similarly, in attitude mechanics we specify a rotation state
of a body by the three Euler angles
$\,q_j\,=\,\phi\,,\,\theta\,,\,\psi\,$ and their momenta
$\,\Phi\,,\;\Theta\,,\;\Psi\,$. After that, we can perform a
canonical transformation to the afore described Serret variables
(which are, in the unperturbed case, merely constants of motion).
A different choice of the generating function would lead one to a
different set of arbitrary constants, one consisting of the
Andoyer variables' initial values: $\,\{\,L_o\,,\,G\,,\,H\,,\,{\it
l}_o\,,\,g_o\,,\,h\,\}\,$, where $\,L_o\,,\,{\it l}_o\,$ and
$\,g_o\,$ are the initial values of $\,L\,,\,{\it l}\,$ and
$\,g\,$. The latter set\footnote{~A similar set consisting of the
initial values of Andoyer-type variables was pioneered by
Fukushima and Ishizaki \cite{fukushimaishizaki}.} (which we shall
call ``modified Andoyer set") consists only of constants of
integration, and hence the corresponding Hamiltonian becomes nil.
Therefore, these constants are the true analogues of the Delaunay
set with $\;M_o\;$ (while the conventional Andoyer set is
analogous to the Delaunay set with $\,M\,$ used instead of
$\,M_o\,$.). The main result obtained below for the modified
Andoyer set $\,\{\,L_o\,,\,G\,,\,H\,,\,{\it
l}_o\,,\,g_o\,,\,h\,\}\,$ will then be easily modified for the
conventional Andoyer set of variables
$\,\{\,L\,,\,G\,,\,H\,,\,{\it l}\,,\,g\,,\,h\,\}\,$.

All in all, the canonical treatment of both orbital and rotational
cases begins with
 \ba
 \dot{\bfq}\;=\;\frac{\partial {\cal H}^{(o)}}{\partial \bfp}\;\;\;,\;\;\;\;\;~~
 \dot{\bfp}\;=\;-\;\frac{\partial {\cal H}^{(o)}}{\partial \bfq}
 ~~~,~~~~~~~~~~~~~~~~~~~
 \label{97}
 \ea
$\bfq$ and $\bfp$ being the coordinates and their conjugated
momenta, in the orbital case, or the Euler angles and their momenta,
in the rotation case. Then one switches, by a canonical
transformation
 \ba
 \nonumber
 \bfq\;=\;f(\bfQ\,,\;\bfP\,,\;t)\;\;\;\\
 \label{98}\\
 \nonumber
 \bfp\;=\;\chi(\bfQ\,,\;\bfP\,,\;t)\;,\;
 \ea
 to
 \ba
 \dot{\bfQ}\;=\;
 \frac{\partial {\cal H}^*}{\partial \bfP}\;=0\;\;\;,\;\;\;\;\;
 \dot{\bfP}\;=\;-\;\frac{\partial {\cal H}^*}{\partial \bfQ}\;=\;0
 \;\;\;,\;\;\;\;
 {\cal H}^*\;=\;0\;\;,
 \label{99}
 \ea
 where $\bf Q$ and $\bf P$ are the Delaunay variables, in the
 orbital case, or the (modified, as explained above) Andoyer
 variables $\;\{\,L_o\,,\;G\,,\;H\,,\;{\it l}_o\,,\;g_o\,,\;h\,\}\;$, in
 the attitude case.

This algorithm is based on the circumstance that an unperturbed
Kepler orbit (and, similarly, an undisturbed Euler
cone) can be fully defined by six parameters so that:\\
 ~\\
 ~~ \textbf{\underline{1.}}~~~These parameters are canonical
variables $\,\{\,\bfQ\,,\,\bfP\,\}\,$ with a zero Hamiltonian:
$\,{\cal H}^*(\bfQ,\,\bfP)\,=\,0\;$; and therefore these parameters are constants. \\
 ~\\
 ~~\textbf{\underline{2.}}~~$\,$For constant $\,\bfQ\,$ and
$\,\bfP\,$, the transformation equations (\ref{98}) are equivalent
to the equations of motion (\ref{97}).

\subsection{The canonical treatment of perturbations}

Under perturbations, the ``constants" $\,\bfQ\,,\,\bfP\,$ begin to
evolve, so that after their substitution into
 \ba
 \nonumber
 \bfq\;=\;f\left(\,\bfQ(t)\,,\;\bfP(t)\,,\;t\,\right)\;\;\;\\
 \label{100}\\
 \nonumber
 \bfp\;=\;\chi(\,\bfQ(t)\,,\;\bfP(t)\,,\;t\,)\;\;\;
 \ea
($f\,$ and $\,\chi\,$ being the same functions as in
(\ref{98})$\,$), the resulting motion obeys the disturbed
equations
  \ba
 \dot{\bfq}\;=\;\frac{\partial \left({\cal H}^{(o)}\,+\,\Delta {\cal H} \right)}{\partial \bfp}
 \;\;\;,\;\;\;\;\;~~
 \dot{\bfp}~=~-~\frac{\partial \left({\cal H}^{(o)}\,+\,\Delta {\cal H} \right)}{\partial \bfq}
 ~~~.~~~~~~~~~~~~~~~
 \label{101}
 \ea
We want our ``constants" $\;\bfQ\;$ and $\;\bfP\;$ also to remain
canonical and to obey
  \ba
 \dot{\bfQ}~=~\frac{\partial \left({\cal H}^*\,+\,\Delta {\cal H}^* \right)}{\partial \bfP}~~~,
 \;\;\;\;\;~~
 \dot{\bfP}\;=\;-\;\frac{\partial \left({\cal H}^*\,+\,\Delta {\cal H}^* \right)}{\partial \bfQ}
 ~~~~~~~~~~~~~~~~~~
 \label{102}
 \ea
 where
 \ba
 {\cal H}^*\,=\;0\;\;\;\;\mbox{and}\;\;\;\;\;\Delta {\cal
 H}^*\left(\bfQ\,,\;\bfP\,\;t\right)\;=\;\Delta {\cal
 H}\left(\,\bfq(\bfQ,\bfP,t)\,,\;\bfp(\bfQ,\bfP,t)\,,\;t\,\right)\;.\;\;
 \label{103}
 \ea
Above all, it is often desired that the perturbed ``constants"
$\,C_j\equiv\,Q_1\,,\,Q_2\,,\,Q_3\,,$ $P_1\,,\,P_2\,,\,P_3\;$ (the
Delaunay constants, in the orbital case, or the modified Andoyer
variables, in the rotation case) be osculating. This demand means
that the perturbed velocity should be expressed by the same function
of $\,C_j(t)\,$ and $\,t\,$ as the unperturbed velocity used to. In
other words, the instantaneous ``simple motions" parameterised by
the ``constants" should be tangent to the perturbed trajectory. (In
the orbital case, this situation is shown on Fig.
\ref{fig:michael2}.) Let us check if osculation is always preserved
under perturbation. The perturbed velocity reads
 \ba
 \dot{\bfq}\;=\;\bfg\;+\;\bfPhi ~~~~~~~~~~~~~~~~~~~~~~~~~~~~~~
 \label{104}
 \ea
where
 \ba
 \bfg(C(t),\,t)\;\equiv\;\frac{\partial \bfq(C(t),\,t)}{\partial t}
 \;\;~~~~~~~~~~~~
 \label{105}
 \ea
is the functional expression for the unperturbed velocity; and
 \ba
 \bfPhi(C(t),\,t)\;\equiv\;\sum_{j=1}^6\,\frac{\partial q(C(t),\,t)}{\partial
 C_j}\;\dot{C}_j(t)\;
 \label{106}
 \ea
is the convective term. Since we chose the ``constants" $\,C_j\,$
to make canonical pairs $\,(\bfQ,\,\bfP)\,$ obeying (\ref{102}) -
(\ref{103}) with vanishing $\,{\cal H}^*\,$, then insertion of
(\ref{102}) into (\ref{106}) will result in
 \ba
 {\bfPhi}\;=\;\sum_{n=1}^3\,\frac{\partial \bfq}{\partial
 Q_n}\;\dot{\bfQ}_n(t)\;+\;\sum_{n=1}^3\,\frac{\partial \bfq}{\partial
 P_n}\;\dot{\bfP}_n(t)\;=\;\frac{\partial \Delta {\cal H}(\bfq,\,\bfp)}{\partial
 \bfp}\;\;\;.
 \label{107}
 \ea
So the canonicity demand is often incompatible with osculation.
Specifically, whenever a momentum-dependent perturbation is present,
we still can use the ansatz (\ref{100}) for calculation of the
coordinates and momenta, but can no longer use (\ref{105}) for
calculating the velocities. Instead, we must use (\ref{104}).
Application of this machinery to the case of orbital motion is
depicted on Fig.\ref{fig:michael1}. Here the constants
$\,C_j=\,(Q_n,\,P_n)\,$ parameterise instantaneous ellipses that,
for nonzero $\bf{\Phi}$, are \emph{not} tangent to the trajectory.
(For details see Efroimsky \& Goldreich \cite{goldreich_2003},
\cite{goldreich_2004}, and Efroimsky \cite{efroimsky2006}.) In the
case of rotational motion, the situation will be identical, except
that, instead of instantaneous Keplerian conics, one will get
instantaneous Eulerian cones (i.e., the loci of the rotational axis,
corresponding to non-perturbed spin states).

 \subsection{From the modified Andoyer variables\\ to the regular ones}

Practical calculations used in the theory of planetary rotation
and in spacecraft attitude dynamics are almost always set out in
terms of the regular Andoyer variables, not in terms of their
initial values (the paper by Fukushima \& Ishizaki
\cite{fukushimaishizaki} being a unique exception). Fortunately,
all our gadgetry, developed above for the modified Andoyer set,
stays applicable for the regular set. To prove this, let us
consider the unperturbed parameterisation of the Euler angles
$\,q_n\,=\,\left(\,\phi\,,\,\theta\,,\,\psi\,\right)\,$ via the
regular Andoyer variables $\,A_j\;=\;(\,{\it
l}\,,\;g\,,\;h\,;\;L\,,\,G\,,\,H\,)\,$:
 \ba
 q_n\;=\;f_n\left(\;A_1(C\,,\,t)\;,\;.\,.\,.\;,\;A_6(C\,,\,t)\;\right)\;\;\;,
 ~~~~~~~~~~~~~~~~~~~~~~
 \label{108}
 \ea
 each element $\,A_i$ being a function of time and of the initial values $\,C_j = ({\it l}_o
 \,,\,g_o\,,\,h\,;\,L_o\,,\,G\,,\,H\,)\,$. The explicit form of parameterisation (\ref{108})
 is given by (\ref{78} - \ref{80}). When a perturbation gets turned on, this parameterisation
 stays, while the time evolution of the elements $\;A_i\;$ changes: beside the standard
 time-dependence inherent in the free-spin Andoyer variables, the perturbed elements acquire
 an extra time-dependence through the evolution of their initial values.\footnote{~This is
 fully analogous to the transition from the unperturbed mean longitude,
 \ba
 \nonumber
 M(t)\;=\;M_o\;+\;n\; \left( \,t\;-\;t_o\, \right)\;\;\;,\;\;\;\;\;{\mbox{with}}\;\;\;\;
 M_o\,,\;n\,,\;t_o\;=\;const\;\;\;,
 \ea
 to the perturbed one,
 \ba
 \nonumber
 M(t)~=~M_o(t)~+~\int_{t_o}^{t}~n(t')~dt'\;\;\;,\;\;\;\;{\mbox{with}}\;\;\;\;
 t_o\,=\;const\;\;\;,\;\;\;\;\;\;\;
 \ea
in orbital dynamics.
 }
 Then the time evolution of an Euler angle $~q_n\,=\,\left(\,\phi\,,~\theta\,,~\psi\,
 \right)\,$ will be given by a sum of two items: (1) the angle's unperturbed dependence upon
 time and time-dependent Andoyer variables; and (2) the convective term $~\Phi_n~$ that arises
 from a perturbation-caused alteration of the Andoyer variables' dependence upon the time:
 \ba
 \dot{q}_n\;=\;\mbox{g}_n\;+\;\Phi_n~~~.~~~~~
 \label{109}
 \ea
 The unperturbed part is
 \ba
 \mbox{g}_n\,=\,\sum_{i=1}^{6}\;\frac{\partial f_n}{\partial A_i}\;
 \left(\frac{\partial A_i}{\partial t}\right)_C\;\;\;,
 \label{110}
 \ea
 while the convective term is given by
 \ba
 \nonumber
 \Phi_n =\sum_{i=1}^{6}\sum_{j=1}^{6}\left(\frac{\partial f_n}{\partial A_i}\right)_t
 \left(\frac{\partial A_i}{\partial C_j}\right)_t\dot{C}_j =
 \sum_{j=1}^{6}\left(\frac{\partial f_n}{\partial
 C_j}\right)_t\,\dot{C}_j~~~~~~~~~~~~~~~
 \ea
 \ba
 =~
 \sum_{j=1}^{3}\left(\frac{\partial f_n}{\partial Q_j}\right)_t\,\dot{Q}_j
 ~+~
 \sum_{j=1}^{3}\left(\frac{\partial f_n}{\partial
 P_j}\right)_t\,\dot{P}_j\,=\;\frac{\partial \Delta{\cal H}(q,p)}{\partial p_n}\;\;,\;
 \label{111}
 \ea
where the set $\;C_j\;$ is split into canonical coordinates and
momenta like this: $\;Q_j\;=\;(\,{\it l}_o\,,\,g_o\,,\,h\,)\;$ and
$\;P_j\;=\;(\,L_o\,,\,G\,,\,H\,)\;$. In the case of free spin they
obey the Hamilton equations with a vanishing Hamiltonian and,
therefore, are all constants. In the case of disturbed spin, their
evolution is governed by (\ref{102} - \ref{103}), substitution
whereof in (\ref{111}) once again takes us to (\ref{107}). This
means that the nonosculation-caused convective corrections to the
velocities stay the same, no matter whether we parameterise the
Euler angles through the modified Andoyer variables (variable
constants) or through the regular Andoyer variables. This invariance
will become obvious if we consider the analogy with orbital
mechanics: in Fig.~\ref{fig:michael1}, the correction $\;\Phibold\;$
is independent of how we choose to parameterise the non-osculating
instantaneous ellipse -- through the set of Delaunay constants
containing $\;M_o\;$, or through the customary set of Delaunay
elements containing $\;M\;$. (The latter set is, historically,
chosen to correspond to a Hamiltonian variation taken with an
opposite sign -- see footnote 8 above. This subtlety, however, is
unimportant to our point.)\\

 \subsection{The Andoyer variables introduced\\
 in a precessing frame of reference}

 \subsubsection{\textbf{Physical motivation}}

 Let us consider a case when the perturbing torque depends not only on the instantaneous
 orientation but also on the instantaneous angular velocity of the rotator. In particular,
 we shall be interested in the fictitious torque emerging when the description is carried out
 in a precessing coordinate system. This situation is often encountered in the theory of
 planetary rotation, where one has to describe a planet's spin not in inertial axes but
 relative to a frame associated with the planet's circumsolar orbital plane (the
 planet's ecliptic). The latter frame is noninertial, because the ecliptic is always
 precessing due to the perturbations exerted by the other planets. The reason why astronomers
 need to describe the planet's rotation not in an inertial frame but in the ecliptic one is
 that this description provides the history of the planet's obliquity, i.e., of the equator's
 inclination on the ecliptic. As the obliquity determines the latitudinal distribution of
 insolation, the long-term history of the obliquity is a key to understanding climate evolution.
 Interestingly, the climate is much more sensitive to the obliquity of the
 planet than to the eccentricity of its orbit. (Murray et al \cite{murray}, Ward
 \cite{ward1973}, \cite{ward1974})

 The canonical theory of rotation of a rigid body in a precessing coordinate frame was
 pioneered by Giacaglia and Jefferys \cite{giacaglia}. It was based on the Andoyer variables
 and was used by the authors to describe rotation of a space station. This theory was greatly
 furthered by Kinoshita \cite{kinoshita77} who applied it to rotation of the rigid Earth.
 Later it was extended by Getino, Ferrandiz and Escapa \cite{escapa_2001},
 \cite{escapa_2002}, \cite{getino} to the case of nonrigid Earth. Simplified versions of the
 Kinoshita theory were employed by Laskar and Robutel \cite{laskar} and by Touma and Wisdom
 \cite{touma93,touma} in their studies of the long-term evolution of the obliquity of Mars. While a
 detailed explanation of this line of research will require a separate review paper, here we
 shall very briefly describe the use of the Andoyer variables in the Kinoshita theory, and
 shall dwell, following Efroimsky \& Escapa \cite{efroimsky_2007}, on the
 consequences of these variables being nonosculating.

 \subsubsection{\textbf{Formalism}}

 Consider an unsupported rigid body whose spin should be described with
 respect to a coordinate system,
 which itself is precessing relative to some inertial axes. The said system is assumed to
 precess at a rate $\,\mubold\,$ so the kinetic energy of rotation, in the inertial axes, is
 given by
 \ba
 T_{kin}\,=\;\frac{1}{2}\;\sum_{i=1}^3\;I_i\,\left(\,\omega_i\,+\;\mu_i\,\right)^2
 \label{112}
 \ea
 where $~\omegabold~$ is the body-frame-related angular velocity of the body relative to the
 precessing coordinate system, while $~\mubold~$ is the angular velocity relative to the
 inertial frame. In (\ref{112}), both $~\omegabold~$ and $~\mubold~$ are resolved into their
 components along the principal axes. The role of canonical coordinates will be played the
 Euler angles $~q_n~=~\phi\,,~\theta\,,~\psi~$ that define the orientation of
 the principal body basis relative to the {\emph{precessing}} coordinate
 basis. To compute their conjugate momenta $~p_n~=~\Phi\,,~
 \Theta\,,~\Psi~$, let us assume that noninertiality of the precessing coordinate system is
 the only angular-velocity-dependent perturbation. Then the momenta are simply the derivatives
 of the kinetic energy. With aid of (\ref{5} - \ref{7}), they can be written as
 \ba
 \Phi=\frac{\partial T_{kin}}{\partial\dot{\phi}}=
     I_1\left(\omega_1+\mu_1\right)\sin \theta \,\sin \psi +
     I_2\left(\omega_2+\mu_2\right)\sin \theta \,\cos \psi +
     I_3\left(\omega_3+\mu_3\right)\cos\theta\;\;,\;\;\;\;\;\;\;
 \label{113}
 \ea
 \ba
 \Theta\;=\;\frac{\partial
 T_{kin}}{\partial\dot{\theta}\;}\;=\;I_3\;\left(\,\omega_3\,+\;\mu_3\,\right)\;\;\;\;,\;\;\;\;\;
 \;\;\;\;\;\;\;\;\;\;\;\;\;\;\;\;\;\;\;\;\;\;\;\;\;\;\;\;\;\;\;\;\;\;\;\;\;\;\;\;\;\;\;\;\;\;\;\;
 \;\;\;\;\;\;\;\;
 \label{114}
 \ea
 \ba
 \Psi\;=\;\frac{\partial T_{kin}}{\partial\dot{\psi}\;}\;=\;
     I_1\;\left(\,\omega_1\,+\;\mu_1\right)\;\cos \psi\; -\;
     I_2\;\left(\,\omega_2\,+\;\mu_2\right)\;\sin
     \psi\;\;\;.~~~~~~~~~~~~~~\,
 \label{115}
 \ea
 These formulae enable one to express the angular-velocity components $~\omega_i~$ and the
 derivatives $~\dot{q}_n\,=\,(\,\dot{\phi}\,,\,\dot{\theta}\,,\,\dot{\psi}\,)~$ via the
 momenta $~{p}_n\,=\,(\,{\Phi}\,,\,{\Theta}\,,\,{\Psi}\,)~$. Insertion of (\ref{113} -
 \ref{115}) into
 \ba
 {\cal H}\,=\,\sum_n\;\dot{q}_n\,p_n \,-\,{\cal L}\,=\,\dot{\phi}\,\Phi\,+\,\dot{\theta}\,\Theta
 \,+\,\dot{\psi}\,\Psi\,-\,T\,+\,V(\phi\,,\;\theta\,,\; \psi)
 \label{116}
 \ea
 results, after some algebra, in
 \ba
 {\cal H}\,=\,T\,+\,\Delta{\cal H}
 \label{117}
 \ea
 where
 \ba
 \nonumber
 \Delta{\cal H}\,=\,-\,\mu_1\,\left[\,\frac{\sin\psi}{\sin\theta}\,\left(\Phi\,-\,\Psi\,\cos
 \theta\right)\,+\,\Theta\,\cos\psi\,\right]                 \\
 \nonumber\\
 \nonumber
 -\,\mu_2\,\left[\,\frac{\cos\psi}{\sin\theta}\,\left(\Phi\,-\,\Psi\,\cos\theta\right)\,-\,\Theta
 \,\sin\psi\,\right]                  \\
 \nonumber\\
                  -\,\mu_3\,\Psi\;+\;V(\phi\,,\;\theta\,,\;\psi)
 ~~~~,~~~~~~~~~~~~~~~~~~\,
 \label{118}
 \ea
 and the potential $\,V\,$ is presumed to depend only upon the angular coordinates,
 not upon the momenta. As can be easily seen from (\ref{25} - \ref{27}), formula
 (\ref{118}) is simply another form of the relation $\,\Delta {\cal{H}}\,=\,-\,
 \mubold\cdot{\mbox{\bf{g}}}\,$ which, according to (\ref{1490} - \ref{1492}), can
 also be expressed via the Andoyer variables solely:
 \ba
 \Delta {\cal{H}}\;=\;-\;\mubold\cdot{\mbox{\bf{g}}}\;=\;-\;\mu_1\;\sqrt{G^2\,-\;L^2}\;\sin{\it{l}}
 -\;\mu_2\;\sqrt{G^2\,-\;L^2}\;\cos{\it{l}}\;-\;\mu_3\;L\;\;\;.\;\;\;\;
 \label{}
 \ea

 Now let us employ the machinery described in the preceding subsection. The
 Euler angles connecting the body axes with the precessing frame will now be
 expressed via the Andoyer variables by means of (\ref{108}). (The explicit
 form of the functional dependence (\ref{108}) is given by (\ref{78} -
 \ref{80}), but this exact form is irrelevant to us.) The fact that the
 Andoyer variables are introduced in a noninertial frame is accounted for by the
 emergence of the $\,\mu$-terms in the expression (\ref{118}) for the
 disturbance $\,\Delta \cal H\,$. Insertion of (\ref{118}) into (\ref{111})
 entails:
 \ba
 \dot{q}_n\,=\,\mbox{g}_n\,+\,\frac{\partial \Delta \cal H}{\partial p_n}
 \label{119}
 \ea
 the convective terms being given by
 \ba
 \frac{\partial \Delta\cal H}{\partial P_1}\;=\;\frac{\partial \Delta \cal H}{\partial
 \Phi}\;=\;\,-\;\,\frac{~\mu_1\;\sin\psi\;+\;\mu_2\;\cos\psi~}{\sin\theta}\;\;\;,~~~~~~~~~~~~~
 \label{120}
 \ea
 \ba
 \frac{\partial \Delta \cal H}{\partial P_2}\;=\;\frac{\partial \Delta \cal H}{\partial
 \Theta}\;=\;\,-\;\,\mu_1\;\cos\psi\;+\;\mu_2\;\sin\psi\;\;\;\;\;\,,~~~~~~~~~~~~~
 \label{121}
 \ea
 \ba
 \frac{\partial \Delta \cal H}{\partial P_3}\;=\;\frac{\partial \Delta \cal H}{\partial
 \Psi}\;=\;\left(\,\mu_1\;\sin\psi\;+\;\mu_2\;\cos\psi\,\right)\;\cot\theta\;-\;\mu_3~~~.\,
 \label{122}
 \ea
 It should be stressed once again that the indices $\;n\,=\,1\,,\,2\,,\,3\;$
 in (\ref{119}) number the Euler angles, so that $\;\dot{q}_n\;$ stand for $
 \,\,\dot{\phi}\,,\,\dot{\theta}\,,\,\dot{\psi}\;$, and $\;{p}_n\;$ signify
 $\,\,{\Phi}\,,\,{\Theta}\,,\,{\Psi}\;$. At the same time, the subscripts
 $\;i\,=\,1\,,\,2\,,\,3\;$ accompanying the components of $\,\mubold\,$ in
 (\ref{120}) correspond to the principal body axes.

 \subsubsection{\textbf{The physical interpretation of the Andoyer variables defined in a
  precessing frame.}}

 The physical content of the Andoyer construction built in inertial axes is
 transparent: by definition, the element $~G~$ is the magnitude of the
 angular-momentum vector, $~L~$ is the projection of the angular-momentum
 vector on the principal axis $~{\bf{\hat{b}}}_3~$ of the body, while $~H~$
 is the projection of the angular-momentum vector on the $~{\bf{\hat{s}}}_3~
 $ axis of the inertial coordinate system. The variable $~h~$ conjugate to
 $~H~$ is the angle from the inertial reference longitude to the ascending
 node of the invariable plane (the one perpendicular to the angular
 momentum). The variable $~g~$ conjugate to $~G~$ is the angle from the
 ascending node of the invariable plane on the reference plane to the
 ascending node of the equator on the invariable plane. Finally, the
 variable conjugate to $~L~$ is the angle $~l~$ from the ascending node of
 the equator on the invariable plane to the the $~{\bf\hat{b}}_1~$ body axis.
 Two auxiliary quantities defined through
 \ba
 \nonumber
 \cos I\;=\;\frac{H}{G}\;\;\;\;,\;\;\;\;\;\;\cos J\;=\;\frac{L}{G}\;\;\;,
 \label{128}
 \ea
 too, have evident physical meaning: $~I~$ is the angle between the
 angular-momentum vector and the $~{\bf\hat{s}}_3~$ space axis, while $~J~$
 is the angle between the angular-momentum vector and the $~{\bf\hat{b}}_3~$
 principal axis of the body, as can be seen on Fig.~\ref{frames}.

 Will all the Andoyer variables and the auxiliary angles $~I~$ and $~J~$
 retain the same physical meaning if we re-introduce the Andoyer
 construction in a noninertial frame? The answer is affirmative, because a
 transition to a noninertial frame is no different from any other
 perturbation: precession of the fiducial frame $~\left({\bf\hat{s}}_1\,,\,
 {{\bf\hat{s}}_2}\,,\,{{\bf\hat{s}}_3}\right)~$ is equivalent to emergence of
 an extra perturbing torque, one generated by the inertial forces. In the
 original Andoyer construction assembled in an inertial space, the
 invariable plane was orthogonal to the instantaneous direction of the
 angular-momentum vector: If the perturbing torques were to instantaneously
 vanish, the angular-momentum vector (and the invariable plane orthogonal
 thereto) would freeze in their positions relative to the fiducial axes $~
 \left({\bf\hat{s}}_1\,,\,{{\bf\hat{s}}_2}\,,\,{{\bf\hat{s}}_3}\right)~$
 (which were inertial and therefore indifferent to vanishing of the
 perturbation). Now, that the Andoyer construction is built in a precessing
 frame, the fiducial plane is no longer inertial. Nevertheless if the
 inertial torques were to instantaneously vanish, then the invariable plane
 would still freeze relative to the fiducial plane (because the fiducial
 plane would cease its precession). Therefore, all the variables will retain
 their initial meaning. In particular, the variables $\;I\;$ and $\;J\;$
 defined as above will be the angles that the angular-momentum makes with
 the precessing $~{\bf\hat{s}}_3~$ space axis and with the $~{\bf\hat{b}}_3~
 $ principal axis of the body, correspondingly.

 \subsubsection{\textbf{Calculation of the angular velocities via the
 Andoyer\\ variables introduced in a precessing frame of reference}}

 Let us now get back to formulae (\ref{5} - \ref{7}) for the principal-body-axes-related
 components of the angular velocity. These formulae give the angular velocity as a function of
 the rates of Euler angle's evolution, so one can symbolically denote the functional dependence
 (\ref{5} - \ref{7}) as $\;\omegabold\,=\,\omegabold(\dot{q})\;$. This dependence is linear,
 so
 \ba
 \omegabold\left(\,\dot{q}(A)\,\right)\;=\;\omegabold(\,\mbox{g}(A)\,)~+~
 \omegabold\left(~{\partial \Delta \cal H}/{\partial p} \;\right)\;\;\;,
 \label{123}
 \ea
 $A\;$ being the set of Andoyer variables. Direct substitution of (\ref{120} - \ref{122})
 into (\ref{5} - \ref{7}) will then show that the second term on the right-hand side in
 (\ref{123}) is exactly $\;\;-\,\mubold\;$:
  \ba
 \omegabold(\dot{q}(A)) \;=\;\omegabold(\mbox{g}(A))\;-\;\mubold\;\;\;.
 \label{124}
 \ea
 Since the ``total" angular velocity $\;\omegabold(\dot{q})\;$ is that of
 the body frame relative to the precessing frame, and since $~\mubold~$ is
 that of the precessing frame relative to some inertial frame, then $\;
 \omegabold(\,\mbox{g}(A)\,)\;$ {\emph{will always return the angular
 velocity of the body relative to the inertial frame of reference, despite
 the fact that the Andoyer variables $\,A\,$ were introduced in a precessing
 frame.}} This nontrivial fact has immediate ramifications for the theory of
 planetary rotation. These will be considered in the subsequent subsections.

  In short, the above may be summarised as:
 \begin{eqnarray}
 \left.
 \begin{array}{ccc}
 \omegabold\left(\,\dot{q}(A)\,\right)\;=\;\omegabold(\,\mbox{g}(A)\,)~+~
 \omegabold\left(~{\partial \Delta \cal H}/{\partial p}
 \;\right)\;\;\;,\;\\
 ~\\
 \omegabold(\,\dot{q}(A)\,)\,=\;\omegabold^{^{(rel)}}\;\;,~~~~~~~~~~~~~~~~~
 ~~~~~~~~~~~~~~~\\
 ~\\
\omegabold(\,\partial \Delta{\cal H}/\partial p\,)\,=\,\mubold
~~~,\,~~~~~~~~~~~~~~~~~~~~~~~~~~~~~~\\
~\\
 \omegabold^{^{(rel)}}\,=\;\omegabold^{^{(inert)}}-\,\mubold\,
 ~~~.~~~~~~~~~~~~~~~~~~~~~~~~~~~
 \end{array}
 \right\}\;\;\;\Longrightarrow\;\;\;~~
 \omegabold(\,\mbox{g}(A)\,)\;=\;\omegabold^{^{(inert)}}\;\;\;\;~~~~~~
 \label{4448}
 \end{eqnarray}
 where the entities are defined as follows:
 \ba
 \nonumber
 \omegabold^{^{(rel)}}\;\equiv\;\; \;\mbox{the relative angular
 velocity,}\,~~~~~~~~~~~~~~~~~~~~~~~~~~~~~~~~~~~~~~~~~~~~~~~~~~~~~\\
 \nonumber
 \mbox{~~~~~~~i.e., the body's angular velocity relative to a
 precessing orbital frame;}
 \nonumber
 \ea
 \ba
 \nonumber
 \mubold\;\,\;\;\;\;\equiv\;\,\;\mbox{the precession rate of that frame with respect to
 an inertial one;~~~~~~~}
 \nonumber
 \ea
 \ba
 \nonumber
 \omegabold^{^{(inert)}}\equiv\;\;\,\mbox{the inertial angular
 velocity,}~~~~~~~~~~~~~~~~~~~~~~~~~~~~~~~~~~~~~~~~~~~~~~~~~~~~~~\\
 \nonumber
 \mbox{~~~~~~~i.e., the body's angular velocity with respect to the inertial frame.~~~~}
 \ea
 To better understand the origin of (\ref{4448}), let us get back to
 the basic Andoyer formalism introduced in the previous subsection. As the first step,
 we introduce, in an unperturbed setting (i.e., for an unsupported rigid
 rotator considered in an inertial frame), the parameterisation of the Euler
 angles $\,q_n\,=\,\left(\,\phi\,,\,\theta\,,\,\psi\,\right)\,$ through the
 Andoyer variables $\,A_j\;=\;(\,{\it l}\,,\;g\,,\;h\,;\;L\,,\,G\,,\,H\,)\,$:
 \ba
 q_n\;=\;f_n\left(\;A_1(C\,,\,t)\;,\;.\,.\,.\;,\;A_6(C\,,\,t)\;\right)\;\;
 \;,~~~~~~~~~~~~~~~~~~~~~~
 \label{125}
 \ea
 each variable $\,A_i$ being dependent upon its initial value $\,C_i$ and
 the time:
 \ba
 A_i\;=\;C_i\;+\;\int_{t_o}^{t} n_i\;dt
 \label{126}
 \ea
 where the mean motions\footnote{~Three of the six Andoyer variables have
 vanishing $~n_i~$ in the unperturbed free-spin case, but at this point it
 is irrelevant.} $\;n_i\;$ may bear dependence upon the other Andoyer
 variables $\;A_j\;$. As already mentioned above, this is analogous to the
 evolution of the mean longitude $\;M\,=\,M_o\,+\,n\,(t\,-\,t_o)\;$ in the
 undisturbed two-body problem (where the mean motion $\;n\;$ is a function
 of another orbital element, the semimajor axis $\;a\;$).

 The initial values $\,C_i\,$ are integration constants, so in the unperturbed case one can
 calculate the velocities $\dot{q}_n=\left(\dot{\phi}\,,\,\dot{\theta}\,,\,\dot{\psi}
 \right)$ simply as the partial derivatives
 \ba
 \mbox{g}_n(A)\,\equiv\,\left(\,\frac{\partial q_n(A(C,\,t))}{\partial t}\,\right)_{_C}\,\equiv\,
 \sum_i\,\frac{\partial q_n(A)}{\partial A_i}
 \left(\,\frac{\partial A_i(C,\,t)}{\partial t}\,\right)_{_C}\;\;\;
 \label{127}
 \ea

 As the second step, we employ this scheme in a perturbed setting. In particular, we introduce a
 perturbation caused by our transition to a coordinate system precessing at a rate $\;\mubold\;$.
 Our Euler angles now describe the body orientation relative to this precessing frame. By
 preserving the parameterisation (\ref{125}), we now introduce the Andoyer variables $\;A_i\;$ in
 this precessing frame. Naturally, the time evolution of $\;A_i\;$ changes because the
 frame-precession-caused Hamiltonian disturbance $\;\Delta \cal H\;$ now shows itself in the
 equations of motion. This disturbance depends not only on the body's orientation but also on its
 angular velocity, and therefore our Andoyer variables cannot be osculating (see equation
 (\ref{107}) and the paragraph thereafter). This means that the unperturbed velocities, i.e., the
 partial derivatives (\ref{127}) no longer return the body's angular velocity relative to the
 precessing frame (i.e., relative to the frame wherein the Andoyer variables were introduced).
 This angular velocity is rather given by the sum (\ref{109}). However, as explained above, the
 unperturbed expression $~\mbox{g}(A)~$, too, has a certain physical meaning: when plugged into $~
 \omegabold(\mbox{g})~$, it always returns the angular velocity {\emph{in the inertial frame}}. It does
 so even despite that now the Andoyer parameterisation is introduced in a precessing
 coordinate frame.\footnote{~This parallels a situation in orbital dynamics, where the
 role of canonical variables is played by the Delaunay constants -- see expression
 (\ref{96}) and a footnote accompanying it.
 In the unperturbed setting (the two-body problem in inertial axes), the
 Cartesian coordinates $\,\erbold\,\equiv\,(\,x_1\,,\,x_2\,,\,x_3)\,$ and velocities $\,(\,
 \dot{x}_1\,,\, \dot{x}_2\,,\,\dot{x}_3)\,$ are expressed via the time and the Delaunay constants
 by means of the following functional dependencies:
 \ba
 \nonumber
 \erbold~=~\efbold\left(\,C\,,~t\,\right)~~~~~\mbox{and}~~~~~{\bf v}~=~
 {\bf g}\left(\,C\,,~t\,
 \right)~~~,\;\;\;\;\mbox{where}\;\;\;{\bf g}\;\equiv\;
 {\partial\efbold}/{\partial t}\;\;\;.
  \ea
 If we want to describe a satellite orbiting a precessing oblate planet, we may fix our reference
 frame on the precessing equator of date. Then the two-body problem will get amended with two
 disturbances. One, $\;\Delta{\cal H}_{oblate}\;$, caused by the presence of the equatorial bulge
 of the planet, will depend only upon the satellite's position. Another one, $\;\Delta{\cal
 H}_{precess}~$, will stem from the noninertial nature of our frame and, thus, will give birth to
 velocity-dependent inertial forces. Under these perturbations, the Delaunay constants will
 become canonical variables evolving in time. As explained in subsection 4.2, the
 velocity-dependence of one of the perturbations involved will make the Delaunay variables
 nonosculating. On the one hand, the expression $~\erbold~=~\efbold
 \left(\,C(t)\,,\;t\,\right)\;$ will return the correct Cartesian coordinates of the satellite in
 the precessing equatorial frame, i.e., in the frame wherein the Delaunay variables were
 introduced. On the other hand, the expression $~{\bf g} \left(\,C\,,\;t\,\right)~$ will no longer
 return the correct velocities in that frame. Indeed, according to (\ref{104} - \ref{107}), the
 Cartesian components of the velocity in the precessing equatorial frame will be given by $~
 {\bf g} \left(\,C\,,\;t\,\right)\;+\;{\partial \Delta{\cal H}_{precess}}/{\partial p}~$. However,
 it turns out that $~{\bf g} \left(\,C\,,~t\,\right)~$ renders the velocity
 with respect to the inertial frame
 of reference. (Efroimsky \cite{efroimsky2006}, \cite{efroimsky2005})}

 \subsubsection{\textbf{Example 1. The theory of Earth rotation}}

 As an example, let us consider the rigid-Earth-rotation theory by Kinoshita
 \cite{kinoshita77}. Kinoshita began with the standard Andoyer formalism in inertial axes. He
 explicitly wrote down the expressions (\ref{108}) for the Euler angles $~q_n\,=\,f_n(A)~$ of
 the figure axis of the Earth, differentiated them to get the expressions for the velocities
 $~\dot{q}_n\,$ as functions of the Andoyer variables: $~\dot{q}_n\,=\,\mbox{g}_n(A)~$, and
 then used those expressions to write down the angles $~I_r(\mbox{g}(A))~$ and $~h_r(\mbox{g}(A))~$
 that define the orientation of the rotation axis.\footnote{~For the Euler angles, Kinoshita
 chose notations, which are often used by astronomers: $\;h\,,\;I\,,\;\phi\;$ and which are
 different from the convention $~q_n\,=\,\phi\,,\,\theta\,,\,\psi~$ used in physics.}
 Kinoshita pointed out that one's knowledge of the solar and lunar torques, exerted on the
 Earth due to its nonsphericity, would enable one to write down the appropriate Hamiltonian
 perturbations and to calculate, by the Hori-Deprit method (\cite{Hori} -
 \cite{Kholshevnikov_1985}), the time evolution of the Andoyer variables. Substitution thereof
 into (\ref{125}) would then give the time evolution of the Euler angles that define the
 figure axis of the planet. Similarly, substitution of the calculated time dependencies of the
 Andoyer variables $\,A\,$ into the expressions for $~I_r(\mbox{g}(A))~$ and $~h_r(\mbox{g}(A))
 ~$ would yield the time evolution of the planet's rotation axis.

 The situation, however, was more complicated because Kinoshita's goal was to calculate the
 dynamics relative to the precessing ecliptic plane. To achieve the goal, Kinoshita amended the
 afore described method by adding, to the lunisolar perturbations, the momentum-dependent
 frame-precession-caused term (see our formula (\ref{118}) above). Stated differently, he
 introduced the Andoyer variables in a precessing frame of reference. This made his Andoyer
 variables $\,A\,$ nonosculating. Kinoshita missed this circumstance and went on to calculate the
 time dependence of the so introduced Andoyer variables. He then plugged those into the
 expressions $\;q_n\,=\,f_n(A)\;$ for the Euler angles of the figure and into the expressions
 $~I_r(\mbox{g}(A))~$ and $~h_r(\mbox{g}(A))~$ for the orientation angles of the Earth rotation
 axis. The expressions $\;q_n\,=\,f_n(A)\;$ still gave him the correct Euler angles of the
 Earth figure (now, relative to the precessing ecliptic plane). The expressions $~I_r(\mbox{g}(A))~$
 and $~h_r(\mbox{g}(A))~$ did $NOT$ give him the correct orientation of the angular-velocity
 vector relative to the precessing frame, because the Andoyer variables introduced by Kinoshita
 in the precessing frame were nonosculating. This peculiar feature of his theory has long
 been ignored, because no direct methods of measurement of the angular velocity of the Earth
 had been developed until 2004. While the thitherto available observations referred only to
 the orientation of the Earth figure (Kinoshita \cite{kinoshita78}), a technique based
 on ring laser gyroscopes (Schreiber et al. \cite{schreiber}) and a VLBI technique
 (Petrov 2007 \cite{petrov})
 later made it possible to directly measure the instantaneous angular velocity of
 the Earth {\it{relative to an inertial frame}}.

 As we demonstrated above, when the Andoyer variables $\;A\;$ are introduced in a precessing
 frame, the expressions $\;q_n\,=\,f_n(A)\;$ return the Euler angles of the body {\it{relative to
 this precessing frame}}, while $\;\omegabold(\mbox{g}_n(A)\,)\;$ returns the
 body-frame-related angular velocity {\it{relative to the inertial frame}}. Accordingly, the
 expressions $~I_r(\mbox{g}(A))~$ and $~h_r(\mbox{g}(A))~$ return the direction angles (as
 seen by an observer located on the body) of the instantaneous angular velocity relative to
 the {\it{inertial frame}}, i.e., the velocity observed in \cite{schreiber}. This way, what
 might have been a problem of the canonical Kinoshita theory became its advantage. It is
 exactly due to the nonosculation of the Andoyer elements, introduced in a precessing
 ecliptic frame, that this theory always returns the angular velocity of the Earth relative to
 inertial axes. We see that sometimes loss of osculation may be an advantage of the theory.

 \subsubsection{\textbf{Example 2. The theory of Mars rotation.}}

 Two groups (\cite{laskar}, \cite{touma93,touma}) independently investigated the long-term evolution
 of Mars' rotation, using a simplified and averaged version of the Kinoshita Hamiltonian. The
 goal was to obtain a long-term history of the Martian spin axis' obliquity, i.e., of the
 angle between the Martian spin axis and a normal to the Martian ecliptic. Just as in the
 afore described case of the Earth, the Martian spin axis is evolving due to the solar torque
 acting on the oblate Mars, while the Martian ecliptic plane is in precession due to the
 perturbations exerted upon Mars by the other planets. Since for realistic rotators the
 Andoyer angle $\,J\,$ is typically very small (i.e., since the angular-velocity and
 angular-momentum vectors are almost parallel), one may, in astronomical applications,
 approximate the obliquity with the angle made by the planet's angular-momentum vector and
 the ecliptic. When the Andoyer construction is built in a precessing frame (the fiducial
 basis $~\left({\bf\hat{s}}_1\,,\,{{\bf\hat{s}}_2}\,,\,{{\bf\hat{s}}_3}\right)~$ being fixed
 on the planet's ecliptic), this assertion means that the obliquity is approximated with
 the Andoyer angle $~I~$. When the Andoyer variables are introduced in the traditional way
 (the basis $~\left({\bf\hat{s}}_1\,,\,{{\bf\hat{s}}_2}\,,\,{{\bf \hat{s}}_3}\right)~$ being
 inertial), the above assertion means that one has to find the orientation of the angular
 momentum relative to the the inertial axes (i.e., to calculate the Euler angles $\;h\;,\;I\;$)
 and to find the orientation of the ecliptic relative to the same inertial axes. Then the
 orientation of the angular momentum with respect to the ecliptic will be found, and it will
 be an approximation for the obliquity.

 The former approach was implemented by Laskar and Robutel \cite{laskar}, the latter by Touma and Wisdom \cite{touma93,touma}. Despite that these two teams introduced the Andoyer variables in different frames of reference,
 the outcomes of their calculations were very close, minor differences being attributed to other reasons.\footnote{~While Touma and Wisdom \cite{touma93,touma} employed unaveraged equations of motion,
 Laskar and Robutel \cite{laskar} used orbit-averaged equations.}
 This coincidence stems from the afore explained fact that the Andoyer
 variables and the Andoyer angles $\;I\,,\,J\,$ retain their
 physical meaning when introduced in a precessing frame. (See subsection 4.4.3.)

\section{The Sadov Variables}

 The Andoyer variables have the merit that they reduce the Hamiltonian
 of an unsupported and torque-free rigid body to one and a half degrees of
 freedom, the resulting expression for the Hamiltonian being very simple.
 However, except for the case of axial symmetry, these variables are not
 action--angle.

 Sadov \cite{sadov} and Kinoshita \cite{kinoshita72}, in an
 independent way, obtained  sets of action--angle variables for the
rotational motion of a triaxial rigid body. Both sets of variables
are very similar. Essentially, they are obtained by solving the
Hamilton-Jacobi equation stemming from the Hamiltonian in the
Andoyer variables. Sadov's transformation is formulated in terms
of the Legendre elliptic functions of the first and third kind,
while in the Kinoshita transformation the Heuman Lambda function
emerges.

 Here we present the approach of Sadov who began by introducing an intermediate set of canonical variables $\,(\beta\,,\;\alpha)\,$ that nullify the Hamiltonian. The generating function ${\cal S}$ of the transformation
\[
(\ell, g, h, L, G, H) \longrightarrow (\beta_1, \beta_2,
\beta_3,\alpha_1, \alpha_2, \alpha_3),
\]
 can be found by solving the corresponding Jacobi equation
 \begin{eqnarray}
 \nonumber
\frac{1}{2} \left(\Frac{\sin^2 \ell}{I_1} + \Frac{\cos^2
\ell}{I_2}\right)
  \left[ \left( \Frac{\partial {\cal S}}{\partial g} \right)^2
   - \left( \Frac{\partial {\cal S}}{\partial \ell} \right)^2\right]
  + \frac{1}{2 I_3} \left( \Frac{\partial {\cal S}}{\partial \ell} \right)^2
  +  \Frac{\partial {\cal S}}{\partial t}  = 0\;\;\;.
  \end{eqnarray}
Since the system is autonomous and variables $g$ and $h$ are
cyclic, the generating function may be expressed as
 \begin{eqnarray}
 \nonumber
{\cal S} = -\alpha_1\, t + \alpha_2 g + \alpha_3 h
     + {\cal U}(\ell; \alpha_1, \alpha_2, \alpha_3) ~~~,
 \end{eqnarray}
wherefrom
\[
\left( \Frac{\partial {\cal U}}{{\partial \ell}} \right)^2 \left[
\left(\frac{1}{I_3}- \frac{1}{I_2}\right) -
\left(\frac{1}{I_1}-\frac{1}{I_2} \right) \sin^2\ell \right] =\left(
2 \alpha_1^2 - \alpha_2^2 \frac{1}{I_2}\right)
  -\alpha_2^2 \left(\frac{1}{I_1}-\frac{1}{I_2} \right)
  \sin^2\ell~~~,
\]
 and further simplification yields a simple equation for the
 function $\,{\cal U}\,$:
 \[
 \left( \Frac{\partial {\cal U}}{{\partial \ell}} \right)^2 =
    I_3\left( \Frac{ a + b \sin^2 \ell}{c + d \sin^2 \ell} \right) ,
 \]
 with
 \[
 \begin{array}{ll}
a = I_1 (\alpha_2^2 - 2 \alpha_1 I_2)\;\;,
\;\;\;& c = I_1 (I_3 - I_2) > 0\;\;,\\[1.5ex]
b = (I_2 - I_1) \alpha_2^2 > 0\;\;,\;\;\;        & d = I_3 (I_2 -
I_1)
> 0\;\;.
 \end{array}
 \]
Hence, the transformation is
\begin{equation}\label{SerTointerm}
\begin{array}{ll}
L = \sqrt{I_3}\;\,\Sqrt{ \Frac{ a + b \sin^2 \ell}{c + d \sin^2
\ell}}\;\;\;,\;\;\; &
   \beta_1 = -t + \Frac{\partial {\cal U}}{{\partial \alpha_1}}\;\;,
   \\ \\ [1.ex]
G = \Frac{\partial {\cal S}}{\partial g} = \alpha_2\;\;,\;\;\; &
   \beta_2 = g + \Frac{\partial {\cal U}}{{\partial \alpha_2}}\;\;, \\ \\ [1.ex]
H = \Frac{\partial {\cal S}}{\partial h} = \alpha_3\;\;,\;\;\; &
   \beta_3 = h\;\;\;,
\end{array}
\end{equation}
whence we see that
\begin{equation}\label{Haminterm}
\mathcal{H} = -\Frac{\partial {\cal S}}{\partial t} = \alpha_1.
\end{equation}

At this point, Sadov introduced\footnote{These quantities already
appear, in a different context, in \cite{tisserand}, p. 394} a
parameter $\kappa$ and a so-called state function $\lambda$ as
\begin{equation}\label{SadovPar}
\kappa^2  = \Frac{I_3(I_2 - I_1)}{I_1(I_3 - I_2)} \geq 0\;\;,\;\;
\qquad
 \lambda^2 = \kappa^2 \,\Frac{I_1}{I_3}\,
               \Frac{2\,I_3\alpha_1 - \alpha_2^2}{\alpha_2^2 - 2\, I_1
                \alpha_1} \geq 0\;\;\;.
\end{equation}
Via these quantities, the angular momentum $L$ in
\Eqref{SerTointerm} may be expressed as
\begin{equation}\label{LSadov}
L^2 = \alpha_2^2 \,\;\Frac{\kappa^2}{\kappa^2 + \lambda^2}\;\,
\Frac{1-\lambda^2 + (\kappa^2 + \lambda^2)\sin^2 \ell}{1 +
\kappa^2 \sin^2 \ell}\;\;\;,
\end{equation}
and the Hamiltonian \Eqref{Haminterm} becomes:
\[
\mathcal{H} =  \Frac{\alpha_2^2}{2\,I_1 I_3}\,
           \Frac{I_3 \lambda^2 + I_1\kappa^2}{\kappa^2 + \lambda^2}.
\]
With this transformation accomplished, Sadov proceeded to a second
one,
\begin{equation}\label{ActioTransf}
(\ell,g,h,L,G,H)\longrightarrow (\varphi_\ell, \varphi_g,
\varphi_h, I_\ell, I_g, I_h)  \;\;\;,
\end{equation}
the new actions being
\[
I_\ell = \Frac{1}{2\pi}\oint L\, d\ell, \qquad I_g =
\Frac{1}{2\pi}\oint G\, d g = G = \alpha_2, \qquad I_h =
\Frac{1}{2\pi}\oint H\, d h = H = \alpha_3\;\;\;.
\]
 While the variables $\,(\beta\,,\,\alpha)\,$ corresponded to a vanishing
 Hamiltonian, the action--angle ones correspond to the initial
 Hamiltonian of Andoyer (though now this Hamiltonian has, of course, to
 be expressed through these new variables).

By means of a convenient auxiliary variable $\,z\,$ introduced
through
\begin{equation}\label{changeAction}
\sin \ell = \Frac{\cos z}{\sqrt{1 + \kappa^2 \sin^2 z}}, \qquad \cos
\ell = -\Frac{\sqrt{1 + \kappa^2}\sin z}{\sqrt{1 + \kappa^2 \sin^2
z}},
\end{equation}
it is possible to derive from \Eqref{LSadov} that
\begin{equation}\label{IellSadov}
I_\ell = \Frac{1}{2\pi}\oint L\, d\ell=
  \Frac{2  \alpha_2 \,\sqrt{1 + \kappa^2}}{\pi\kappa \sqrt{\kappa^2 + \lambda^2}}
  \left[ (\kappa^2 + \lambda^2)  \Pi(\kappa^2, \lambda)
    -  \lambda^2 \, K (\lambda)\right].
\end{equation}
where $K$ and $\Pi$ are the complete elliptical integrals of the
first and the third kind, respectively.

Symbolically, the above expression may be written as $\,I_\ell \,=\,
I_\ell (I_g\,,\, \lambda)\, =\, I_g\, f(\lambda)\,$. According to
the Implicit Function Theorem, there exists a function
$\,\phi(I_\ell / I_g)\,$ inverse, locally, to $\,f(\lambda)\,$. Let
us denote this function as $\,\lambda \,= \,\phi(I_\ell / I_g)\,$.
Sadov \cite{sadov} proved that it is defined for all values
$\,(I_\ell / I_g) \ge0\,$, and that it is analytic for $\,(I_\ell /
I_g) > 0 \,$ or $\,(I_\ell / I_g) \ne (2/\pi) \arctan{\kappa}\,$ at
$\,\lambda =1\,$.

Since the Hamiltonian is known to be
\begin{equation}\label{SadovH}
\mathcal{H} =  \Frac{I_g^2}{2\,I_1 I_3}\,
           \Frac{I_3 \lambda^2 + I_1\kappa^2}{\kappa^2 + \lambda^2},
\end{equation}
the angular variables could be found by integrating the Hamilton
equations
\[
\dot{\varphi}_\ell = \Frac{\partial \mathcal{H}}{\partial
I_\ell}\;, \qquad \dot{\varphi}_g = \Frac{\partial
\mathcal{H}}{\partial I_g}\;, \qquad \dot{\varphi}_h =
\Frac{\partial \mathcal{H}}{\partial I_h}\;\;,
\]
if the explicit relation between the action momenta and $\lambda$
were known. Unfortunately we do not have such a relation at hand. To
circumvent this difficulty, the generating function of the
transformation must be derived.

Let  ${\cal W}$ be the generating function of the transformation
\Eqref{ActioTransf} from the Andoyer variables to the action and
angle variables. It may be chosen as
\[
{\cal W} = I_g g + I_h h + {\cal V}(\ell; I_\ell, I_g)\;\;\;,
\]
and the equations of this transformation will read:
\[
\begin{array}{l@{\qquad}l}
L = \Frac{\partial {\cal V}}{{\partial \ell}}\;\;\;, &
   \varphi_\ell = \Frac{\partial {\cal V}}{{\partial I_\ell}}\;\;\;, \\[1.5ex]
G = \Frac{\partial {\cal W}}{\partial g} = I_g\;\;\;, &
   \varphi_g = \Frac{\partial {\cal W}}{{\partial I_g}} =
         g + \Frac{\partial {\cal V}}{\partial I_g}\;\;\;, \\[1.5ex]
H = \Frac{\partial {\cal W}}{\partial h} = I_h\;\;\;, &
   \varphi_g = \Frac{\partial {\cal W}}{{\partial I_h}} = h\;\;\;.
\end{array}
\]
The generating function ${\cal V}$ is obtained directly from the
quadrature
\[
{\cal V}(\ell; I_\ell, I_g) = \int_{\ell_0}^\ell L(\ell; I_\ell,
I_g)\, d\ell\;\;\;,
\]
where we have to replace $L$ with expression \Eqref{LSadov}.
\par
To calculate $\varphi_\ell$ and $\varphi_g\,$, we need to find the
partial derivatives $\,\partial \lambda / \partial I_\ell\,$ and
$\,\partial \lambda /
\partial I_g\,$. Since
 $\lambda = \phi(I_\ell / I_g) $ and $I_\ell = I_g f(\lambda)\,$,
 then
\[
\Frac{\partial I_\ell}{{\partial \lambda}}
   = I_g \Frac{\partial f}{{\partial \lambda}}
   = \Frac{1}{2\pi} \oint \Frac{\partial L}{{\partial \lambda}} =
 - \Frac{I_g 2 \kappa \lambda (1 + \kappa^2)^{1/2}}{\pi( \kappa^2+\lambda^2)^{3/2}}\,
   K(\lambda)
\]

 After introducing notation $u \equiv I_\ell / I_g\,$, we can write:
\begin{equation}\label{partialPhiU}
\Frac{\partial \phi}{\partial u} = \Frac{1}{\partial u / \partial
\phi} =
  I_g \Frac{1}{\partial I_\ell / \partial \phi} =
  I_g \Frac{1}{\partial I_\ell / \partial \lambda} =
  - \Frac{\pi( \kappa^2+\lambda^2)^{3/2}}{ 2 \kappa \lambda
  (1 + \kappa^2)^{1/2}K(\lambda)}
\end{equation}
 whence
\[
\Frac{\partial \lambda}{\partial I_g} = \Frac{\partial
\phi}{\partial u}\Frac{\partial u}{\partial I_g} =
\Frac{I_\ell}{I_g^2} \Frac{\pi( \kappa^2+\lambda^2)^{3/2}}{ 2 \kappa
\lambda (1 + \kappa^2)^{1/2}K(\lambda)} =
\Frac{(\kappa^2+\lambda^2)}{I_g  \kappa^2 \lambda  K(\lambda)}
\left[(\kappa^2+\lambda^2) \Pi(\kappa^2, \lambda) - \lambda^2
K(\lambda)\right].
\]
Now, taking into account \Eqref{LSadov} and employing the chain
rule, we arrive at
\begin{equation}\label{phiSadovProv}
\begin{array}{l}
\varphi_\ell =\Frac{\partial {\cal V}}{{\partial I_\ell}} =
  \Frac{\partial {\cal V}}{{\partial \lambda}}
     \Frac{\partial \lambda}{{\partial I_\ell}} =
  \Frac{\partial \lambda}{{\partial I_\ell}}
  \Int_{\ell_0}^\ell \Frac{\partial L}{{\partial \lambda}} d\ell\;\;,
  \\[2ex]
\varphi_g =g + \Frac{\partial }{{\partial I_g}}
  \Int_{\ell_0}^\ell  L\, d\ell
  = g + \Frac{1}{I_g}  \Int_{\ell_0}^\ell  L\, d\ell +
  \Frac{\partial \lambda}{{\partial I_g}}
     \Int_{\ell_0}^\ell \Frac{\partial L}{{\partial \lambda}}\, d\ell\;\;\;.
\end{array}
\end{equation}

Straightforward differentiation of \Eqref{LSadov} yields:
\[
\Frac{\partial L}{\partial \lambda} = - \Frac{I_g \kappa \lambda
(1+\kappa^2)}
  {(\kappa^2 + \lambda^2)^{3/2}} \Frac{1}{\sqrt{1 + \kappa^2 \sin^2 \ell}
  \sqrt{1-\lambda^2 + (\kappa^2 + \lambda^2)\sin^2 \ell}}.
 \]

With the help of \Eqref{changeAction}, we can now compute the
quadratures
\[
\Int_{\ell_0}^\ell \Frac{\partial L}{{\partial \lambda}}\, d\ell =
 - \Frac{I_g  \kappa \lambda (1 + \kappa^2)^{1/2}}{( \kappa^2+\lambda^2)^{3/2}}\,
    F(z, \lambda)
\]
and
\[
\Int_{\ell_0}^\ell L\, d\ell =
  \Frac{I_g}{\kappa}\Sqrt{\Frac{1 + \kappa^2}{\kappa^2+\lambda^2}}
   \Big[(\kappa^2+\lambda^2) \Pi(z,\kappa^2, \lambda) - \lambda^2
   K(\lambda)\Big]\;\;\;,
\]

 insertion whereof into Eqs. \Eqref{phiSadovProv} results in
 \begin{equation}
 \label{phiSadov}
\begin{array}{l}
\varphi_\ell = \Frac{\pi}{2} \Frac{F(z,\lambda)}{K(\lambda)},\\[2ex]
\varphi_g  = g + \Frac{1}{\kappa} \Sqrt{(\kappa^2+\lambda^2)(1+
\kappa^2)}
 \left[  \Pi(z,\kappa^2, \lambda)
   - \Frac{ \Pi(\kappa^2, \lambda)F(z,\lambda)}{K(\lambda)} \right]\;\;\;.
\end{array}
\end{equation}

 Inversion of the first of these expressions entails
\[
z = \am (2 \varphi_\ell K(\lambda)/ \pi , \, \lambda )\;\;\;.
\]
(For the definition of the elliptic function $\,am\,$ see Appendix
A.2 below. ) Thus, we have the Andoyer angle $g$ expressed via the
action-angle variables:
\[
g = \varphi_g + \Frac{1}{\kappa} \Sqrt{(\kappa^2+\lambda^2)(1+
\kappa^2)}
 \left[  \frac{2}{\pi}\Pi(\kappa^2, \lambda) \,  \varphi_\ell
 -  \Pi\Big(\am (2 \varphi_\ell K(\lambda)/ \pi\,, \, \lambda ),\kappa^2, \lambda\Big)
 \right]\;\;\;,
\]
 while the angle $\ell$ is obtained directly from expression
\Eqref{changeAction}.

\section{CONCLUSIONS}

In this paper, we have reviewed the Serret-Andoyer (SA) formalism
for modeling and control of rigid-body dynamics from the dynamical
systems perspective. We have dwelt upon the important topic of
modeling of general, possibly angular velocity-dependant
disturbing torques, and upon the inter-connection between the
Andoyer and the Sadov sets of variables. We have also contributed
some new insights.

The first insight is that the Andoyer variables turn out to be
non-osculating in the general case of angular-velocity-dependent
perturbation. The second insight is that even when these variables
are introduced in a precessing reference frame, they preserve
their interconnection with the components of the angular momentum
-- a circumstance that makes the Andoyer variables especially
valuable in astronomical calculations.

In summary, this treatise constitutes a first step towards
understanding the consequences of using the SA formalism as a
single, generic language for modelling rigid body dynamics in
diverse
 -- and seemingly unrelated -- areas such as celestial mechanics, satellite attitude
 control, and geometric mechanics.



 ~\\

 \noindent
{\underline{\bf{\Large{\textbf{Appendix.}}}}} \\
 ~\\

\noindent {\large{\textbf{A.1.~~~Spherical-trigonometry formula (\ref{50})}}}\\

 \noindent
 The standard formula of spherical trigonometry \cite{smart},
 \ba
 \cos g~=~\cos(\phi - h)\,\cos(\psi - l)\,+\,\sin(\phi - h)\,\sin(\psi - l)\,\cos(\pi-\theta)~~,
 ~~~~~~~~~~~~~~~~~~~
 \label{A1}
 \ea
 immediately entails:
 \ba
 \nonumber
 \sin g\;dg\;=\;[\,\sin(\phi - h)\;\cos (\psi - l)\;-\;\sin(\phi - h)\;\sin
 (\psi - l)\;\cos \theta \,]\,d(\phi - h)~~~~~~~~~~~\\
 \nonumber\\
 \nonumber
 +\;[\,\cos(\phi - h)\;\sin (\psi - l)\;-\;\sin(\phi -
 h)\;\cos (\psi - l)\;\cos \theta \,]\,d(\psi - l)~~~~~~~~~~~\\
 \nonumber\\
 +\;\,\sin(\phi - l)\;\,\sin(\psi -
 l)\;\,\sin\theta\;\;d\theta\;\;\,.~~~~~~~~~~~~~~~~
 \label{A2}
 \ea
 The other standard formulae of the spherical trigonometry enable one to transform (\ref{A2})
 into
 \ba
 \nonumber
 \sin g\;dg\;=~~~~~~~~~~~~~~~~~~~~~~~~~~~~~~~~~~~~~~~~~~~~~~~~~~~~~~~~~~~~~~~~~~~~~~~~~~~~~~~~
 ~~~~~~~~~~~~~\\
 \nonumber\\
 \nonumber
 \sin g\;
 \cos I\;d(\phi - h)\;+\;\sin g\;\cos J\;d(\psi - l)\;+\;\sin (\phi - h)\;\sin(\psi - l)\;\sin
 \theta\;d\theta~=~~~
 \ea
 \ba
 \sin g\;\cos I\;d(\phi - h)\;+\;\sin g\;\cos J\;d(\psi -l)\;+\;\sin(\psi -l)
 \;\sin J\;\sin g\;d\theta\;\;\;,~~~~~~~~~~~~~
 \label{A3}
 \ea
 wherefrom equality (\ref{50}) follows.

 ~\\

 \noindent {\large{\textbf{A.2.~~~The dimensionless time $~u~$ and its interconnection\\ $\left.~
 ~~~~~~\right.$ with $~T_{kin}\,,~\,G\,,~\,G\,\cos I\,,~\,\theta\,,~\mbox{and}~\psi~$}}}~\\

 First let us recall some basics. As agreed in the text, we choose the body axes to coincide with
 the principal axes of inertia, $\;{\bf\hat b}_1\,,\,{\bf\hat b}_2\,,\,{\bf\hat b}_3\;$. This
 makes the inertia tensor look like:
 \begin{equation}
 \mathbb{I} =
 \left(
 \begin{array}{ccc}
 I_1 & 0 & 0 \\
 0 & I_2 & 0 \\
 0 & 0 & I_3 \\
 \end{array}
 \right)\;\;\;,
 \label{A4}
 \end{equation}
 Decomposing the body angular velocity $\;\omegabold\;$ over the principal basis,
 \begin{eqnarray}
 \omegabold~=~\omega_1~{\bf{\hat b}}_1~+~\omega_2~{\bf{\hat b}}_2~+~\omega_3~{\bf{\hat b}}_3~~~,
 \label{A5}
 \end{eqnarray}
 we can write twice the kinetic energy as
 \begin{equation}
 2\;T_{kin}\;=\;\omegabold\;{\mathbb{I}}\;\omegabold\;=\;I_1 \omega_1^{2} + I_2 \omega_2^{2} +
 I_3 \omega_3^{2}\;\;\;,
 \label{A6}
 \end{equation}
 and the body-frame-related angular-momentum vector as
  \begin{equation}
  {\bf{g}} = \mathbb{I} \cdot {\omegabold}\;\;\;.
  \label{A7}
  \end{equation}
 The square of this vector will be:
 \begin{equation}
 {\bf{g}}^{2}\;=\;I_1^{2}\;\omega_1^{2}\;+
 \;I_2^{2}\;\omega_2^{2} + I_3^{2} \; \omega_3^{2}\;\;\;,
 \label{A8}
 \end{equation}
 while its direction cosines with respect to an invariable plane-based coordinate system will
 read:\footnote{~The direction cosines can be expressed in terms of the angle $~l~$ from the
 ascending node of the equator on the invariable plane to the the $~{\bf{\hat b}}_1~$ body axis,
 and the angle $~J~$ between the angular-momentum vector and the $ ~{\bf{\hat b}}_3~ $ principal
 axis of the body (Fig.~\ref{frames}):
 \ba
 \nonumber
 \alpha' ~=~ \sin J ~ \sin l~~~,~~~~~~ {\beta}^{'} ~=~ \sin J ~ \cos l~~~~~~ {\gamma}^{'}~=~ \cos
 J~~~,
 \ea
 so that
 \ba
 \nonumber
 I_1\,\omega_1\;=\;G\;\sin J \; \sin l  \;\;\;,\;\;\;\;\;
 I_2\,\omega_2\;=\;G\;\sin J \; \cos l  \;\;\;,\;\;\;\;\;
 I_3\,\omega_3\;=\;G\;\cos J \;\;\;.\;\;\;\;\;
 \ea
 }
 \begin{eqnarray}
 \alpha'\,=\,\frac{\mbox{g}_{1}}{{G}} ~=~ \frac{I_1\;\omega_1}{{G}}\;\;\;,\;\;\;\;\;\;
 {\beta}^{'} \,=\,\frac{\mbox{g}_{2}}{{G}} ~=~ \frac{I_2\;\omega_2}{{G}}\;\;\;,\;\;\;\;\;\;
 {\gamma}^{'}\,=\,\frac{\mbox{g}_{3}}{{G}} ~=~ \frac{I_3\;\omega_3}{{G}}\;\;\;,\;\;\;
 \label{A9}
 \end{eqnarray}
 $G\;$ denoting the magnitude of the angular momentum vector: $\;G\,\equiv\,|{\bf{g}}|\;$.
 At this point it is convenient to define an auxiliary quantity $\;P^2\;$ via
 \begin{eqnarray}
 P^{2}~\equiv~\frac{(\alpha')^{2}}{I_1}~+~\frac{({\beta}^{'})^{2}}{I_2}~+~\frac{
 ({\gamma}^{'})^{2}}{I_3}\;\;\;\;.
 \label{A10}
 \end{eqnarray}
 Substitution of (\ref{A9}) into (\ref{A10}) shows that this auxiliary
 quantity obeys
 \begin{eqnarray}
 \label{K_squared_P_squared}
 G^{2}\;P^{2}\;
 =\;I_1\;\omega_1^{2}\;+\;I_2\;\omega_2^{2}\;+\;I_3\;\omega_3^{2}\;\;\;
 \label{A11}
 \end{eqnarray}
 or, equivalently,
 \begin{eqnarray}
 \label{H_K_P}
 G^{2}\;P^{2}\;=\;2\;T_{kin}\;\;\;\;.
 \label{A12}
 \end{eqnarray}
 We see that $\;P^2\;$ is an integral of motion, a circumstance that will later help us with
 reduction of the problem.

 Due to the evident identity
 \begin{eqnarray}
 \label{dir_cos_identity}
 (\alpha')^{2} + ({\beta}^{'})^{2} + ({\gamma}^{'})^{2} = 1\;\;\;
 \label{A13}
 \end{eqnarray}
 our $\;P^2\;$ depends upon only two directional cosines. Elimination of $\;{\gamma}^{'}\;$ from
 (\ref{A10}), by means of (\ref{A13}), trivially yields:
 \begin{equation}
 P^{2} - \frac{1}{I_3}= \left(\frac{1}{I_1} -  \frac{1}{I_3}\right) (\alpha')^{2} +
 \left( \frac{1}{I_2} -  \frac{1}{I_3}\right) ({\beta}^{'})^{2}.
 \label{A14}
 \end{equation}
 Note that, if we now introduce an $a^{2}$ and a $b^{2}$ such that
 \begin{equation}
 a^{2}~=~\frac{P^{2}~-~\frac{\textstyle 1}{\textstyle I_3}}{\frac{\textstyle 1}{\textstyle I_1}~
 -~\frac{\textstyle 1}{\textstyle I_3}}\;\;\;,
 \label{A15}
 \end{equation}
 \begin{equation}
 b^{2} ~=~ \frac{P^{2} ~-~ \frac{\textstyle 1}{\textstyle I_3}}{\frac{\textstyle 1}{
 \textstyle I_2}~ - ~\frac{\textstyle 1}{\textstyle I_3}}\;\;\;,
 \label{A16}
 \end{equation}
 we shall be able to write down the definition of $\;P^2\;$ in a form that will, formally, be
 identical to definition of an ellipse:
 \begin{equation}
 \frac{(\alpha')^{2}}{a^{2}} + \frac{({\beta}^{'})^{2}}{b^{2}} = 1\;\;\;.
 \label{A17}
 \end{equation}
 This means that, along with $~P^2~$, it is convenient to introduce another auxiliary variable,
 $~\xi~$, one that obeys
 \begin{eqnarray}
 \frac{(\alpha')^{2}}{a^{2}}~ \equiv ~\cos^{2}\xi\;\;\;,\;\;\;\;\;\;\;
 \frac{({\beta }^{'})^{2}}{b^{2}}~ \equiv ~\sin^{2}\xi\;\;\;.
 \label{A18}
 \end{eqnarray}
 Insertion of the above into (\ref{A13}) entails
 \ba
 \label{gamma_squared}
 ({\gamma}^{'})^{2} \;=\;\left(1\; -\; a^{2}\right)\left(1 \;- \;\frac{b^{2} \;-\; a^{2}}{1 \;-\; a^{2}}
 \;\sin^{2}\xi\right)\;\;\;.
 \label{A19}
 \ea
 Then, after defining the quantity $\;\kappa^2\;$ and the function $\;\Delta\;$ as
  \ba
 \label{kappa_squared}
 \kappa^{2}\;\equiv\;  \frac{b^{2}\; -\; a^{2}}{1\; -\; a^{2}}
 \label{A20}
 \ea
 and
 \ba
 \label{delta_xi}
 \Delta \xi\;\equiv\; \sqrt{1 \;-\; \kappa^{2}\;\sin^{2}\xi}\;\;\;,
 \label{A21}
 \ea
 we shall be able to cast the direction cosines of the angular-momentum vector
 in the form of
 \ba
 \label{alpha}
 \alpha' = a\; \cos \xi\;\;\;,\;\;\;\,\;\;
 \label{A22}
 \ea
 \ba
 \label{beta}
 {\beta}^{'} = b\; \sin \xi\;\;\;,\;\;\;\;\;\;\,
 \label{A23}
 \ea
 \ba
 \label{gamma}
 {\gamma}^{'} = \sqrt{1 - a^{2}}\;\Delta \xi\;\;\;.
 \label{A24}
 \ea
 Looking back at (\ref{A9}), we see that initially we started out with three
 direction cosines, only two of which were independent due to the equality
 (\ref{A13}). The latter meant that all these cosines might be expressed via
 two independent variables. The quantities $~P^2~$ and $~\xi ~$ were cast for
 the part. (Mind that $~a~$ and $~b~$ depend on $~P^2~$ through (\ref{A15})
 - (\ref{A16}).) It should also be mentioned that, due to (\ref{A13}), $\;P^2\;$
 is a constant of motion, and therefore formulae (\ref{A22} - \ref{A24})
 effectively reduce the problem to one variable, $~\xi\;$, which thereby plays
 the role of re-scaled time, in terms whereof the problem is fully solved. Below
 we shall explicitly write down the dependence of the ``time" $~\xi~$ upon
 the real time $\;t\;$ or, equivalently, upon the dimensionless time
 $\;u\;\equiv\;n\,(t\,-\,t_o) \;$ emerging in (\ref{54} - \ref{55}).

 Undisturbed spin of an unsupported rigid body obeys the Euler equations for the
 principal-axes-related components of the body angular velocity:
 \ba
 \label{I_1_dp_dt}
  I_1 \;\frac{d\omega_1}{dt} \;=\; \omega_2 \; \omega_3 \;(I_2 \;-\; I_3)\;\;\;,
 \label{A25}
 \ea
 \ba
 \label{I_2_dq_dt}
  I_2 \;\frac{d\omega_2}{dt} \;=\; \omega_3 \; \omega_1 \;(I_3 \;-\; I_1)\;\;\;,
 \label{A26}
 \ea
 \ba
 \label{I_3_dr_dt}
  I_3 \;\frac{d\omega_3}{dt} \;=\; \omega_1 \; \omega_2 \;(I_1 \;-\; I_2)\;\;\;.
 \label{A27}
 \ea
 Insertion of expressions (\ref{A9}) into (\ref{I_1_dp_dt}), (\ref{I_2_dq_dt}), and
 (\ref{I_3_dr_dt}) provides an equivalent description written in terms of the
 angular-momentum
 vector's directional cosines relative to the principal axes:
 \ba
 \label{dalpha_dt}
  \frac{d\alpha'}{dt}\;=\;G\;{\beta}^{'}\;{\gamma}^{'}\;\left(\frac{1}{I_3}\;-\;
  \frac{1}{I_2}\right)~~~,
 \label{A28}
 \ea
 \ba
 \label{dbeta_dt}
 \frac{d{\beta}^{'}}{dt}\;=\;G\;{\gamma}^{'}\;\alpha'\;\left(\frac{1}{I_1}\;-\;\frac{1}{I_3}\right)~~~,
 \label{A29}
 \ea
 \ba
 \label{dgamma_dt}
 \frac{d{\gamma}^{'}}{dt}\;=\;G\;\alpha'\;{\beta}^{'}\;\left(\frac{1}{I_2}\;-\;\frac{1}{I_1}\right)~~~.
 \label{A30}
 \ea
 Substitution of (\ref{alpha} - \ref{gamma}) into (\ref{A28}) will then entail
 \ba
 \label{dxi_over_delta_xi_1}
 \frac{d \xi}{\Delta \xi}~=~\frac{b}{a}~\sqrt{1~-~a^{2}}\,\left(\frac{1}{I_2}~-~\frac{1}{I_3}
 \right)~G~dt\;\;\;.
 \label{A33}
 \ea
 A subsequent insertion of (\ref{A15} - \ref{A16}) and (\ref{A21}) into
 (\ref{A33}) will then entail:
 \ba
 \label{dxi_over_delta_xi_2}
 \frac{d \xi}{\sqrt{1~-~\kappa^{2}~\sin^{2}\xi}}~=~G~dt~\sqrt{\left(\frac{1}{I_1}~-~P^{2}
 \right)\left(\frac{1}{I_2}~-~ \frac{1}{I_3}\right)} \;\;\;.
 \label{A34}
 \ea
 Now define a ``mean motion" $\;n\;$ as
 \ba
 \label{n}
 n~\equiv~G~\sqrt{\left(\frac{1}{I_1}~-~P^{2}\right)~\left(\frac{1}{I_2}~-~\frac{1}{I_3}
 \right)}~~~,
 \label{A35}
 \ea
 Since, according to (\ref{A12}), $~P^2~$ is integral of motion, then so is $~n~$, and
 therefore
 \ba
 \label{u_definition}
  \int_{0}^{\xi} \frac{d\xi}{\sqrt{1~-~\kappa^{2}\;\sin^{2}\xi}}\;=\;\int_{t_o}^{t}\;n\;dt'~=~
  n \;(t\; -\;t_o )\;\;\;.
 \label{A36}
 \ea
This means that, if we define a dimensionless time as
 \ba
 u\;\equiv\;\int_{t_o}^{t}n\;dt'\;=\;n\;(t\;-\;t_o)
 \label{A37}
 \ea
and a function $\;F\;$ as
 \ba
 F(\xi\,,~\kappa)~\equiv~\int_{0}^{\xi} \frac{d\xi'}{\sqrt{1~-~\kappa^{2}~\sin^{2}\xi'}}~~~,
 \label{A38}
 \ea
 then the interrelation between the parameter $\;\xi\;$ and the time will look like:
 \ba
 u\;=\;F(\xi\,,~\kappa)\;\;\;.
 \label{A39}
 \ea
 This is a Jacobi elliptic equation whose solutions are be written in terms of the following
 elliptic functions:
 \ba
 \label{phi}
 \xi ~=~ {\mbox{am}} (u\,,~\kappa )~\equiv~ F^{-1} (u\,,~\kappa ) \;\;\;,
 \label{A40}
 \ea
 \ba
 \label{sin_xi}
 \sin  \xi  ~=~ {\mbox{sn}} (u\,,\;\kappa) ~\equiv~ \sin (\,\mbox{am} (u\,,\;\kappa)\,)\;\;\;,
 \label{A41}
 \ea
 \ba
 \label{cos_xi}
 \cos  \xi  ~=~ {\mbox{cn}} (u\,,\;\kappa) ~\equiv~ \cos (\,\mbox{am} (u\,,\;\kappa)\,)\;\;\;,
 \label{A42}
 \ea
 \ba
 \label{delta_xi_u}
 \Delta\xi~\equiv~\sqrt{1~-~\kappa^2~{\mbox{sn}}^2 (u\,,~\kappa)}~\equiv~{\mbox{dn}}(u\,,~\kappa)
 ~~~.
 \label{A43}
 \ea
 According to (\ref{A39}), $~u~$ is a function of $~\xi~$ and $~\kappa~$. The
 integral of motion $~\kappa~$ is, through (\ref{A20}) and (\ref{A12}), a function
 of $~T_{kin}~$ and $~G~ $, while the parameter $~\xi~$ is, through the medium of
 (\ref{A15} - \ref{A16}) and (\ref{A22} - \ref{A23}), a function of $~T_{kin}~$,
 $~G~$, $~\alpha'\,=\,\sin J~\sin l~$, and $~{\beta}^{'}~=~ \sin J~\cos l~$. All
 in all, $~u~$ is a function of $~T_{kin}~$, $~G~$, $~J~$, and $~l~$.

 It is also possible to express $~u~$ through another set of geometric variables.
 Recall that $~(\,\phi\,,\,\theta\,,\,\psi\,)~$ are the Euler angles defining
 orientation of the body axes $~(\,{\bf\hat{b}}_1\,,\,{\bf\hat{b}}_2\,,\,
 {\bf\hat{b}}_3\,)~$ relative to the fiducial frame $~(\,{\bf\hat{s}}_1\,,\,
 {\bf\hat{s}}_2\,,\,{\bf\hat{s}}_3\,)~$, while $~(\,\phi_{0}\,,\,J\,,\,l\,)~$ are
 the Euler angles defining orientation of the body axes with respect to the
 invariable plane, as on Fig.~2. Out of that picture, it is convenient
 to single out a spherical triangle whose sides are given by $~(\,\phi\,-\,h\,,\,
 \psi\,-\,l\,,\,g\,)~$,
 and whose internal angles are $~(\,I\,,\,\pi\,-\,\theta\,,\,J\,)~$, as shown in
 Fig.~3. An analogue to the law of cosines, for spherical triangles,
 looks like:
 \begin{eqnarray}
 \cos I~=~\cos \theta~\cos J~+~\sin \theta~\sin J~\cos (\psi\,-\,l)~~~,
 \label{A44}
 \end{eqnarray}
 which is the same as (\ref{A1}), up to a cyclic transposition.
 The second term on the right-hand side of (\ref{A44}) can be
 expanded as
 \begin{align}
 \sin J~\cos (\psi\,-\,l)~
   &=\;\sin J\;\left(\,\cos\psi\;\cos l\,+\,\sin\psi\;\sin l \,\right)
  \nonumber \\
   &=\;\left(\,\sin J\;\sin l\,\right)\;\sin\psi ~+~
   \left(\,\sin J ~\cos l\,\right)~\cos \psi
  \nonumber \\
   &=\;{ \alpha }' ~ \sin \psi + { \beta }^{ ' } ~ \cos \psi
  \nonumber \\
   &=\;a~\sin \psi~ \cos (\,am(u)\,) ~+~ b ~\cos \psi ~\sin
   (\,am(u)\,) ~~~,
 \label{A45}
 \end{align}
 where the last line was obtained with aid of (\ref{A22} - \ref{A23}) and
 (\ref{A40}). Thus (\ref{A44}) acquires the shape of
 \begin{eqnarray}
   \cos I~=~\cos\theta~\sqrt{1\,-\,a^{2}}~\Delta\mbox{am}(u)~
   &+&~a~\sin\theta~\sin\psi~\cos( \mbox{am}(u))~
  \nonumber \\
   &+&~b~\cos\psi~\sin(\mbox{am}(u))~~~.
 \label{A46}
 \end{eqnarray}
We see that $\;u\;$ depends upon $\;\psi\,,\;\theta\,,\;\cos I\;$,
and (through $\;a\;$ and $\;b\;$) upon $\;T_{kin}\;$ and $\;G\;$.
This is equivalent to saying that $\;u\;$ is a function of
$\;T_{kin}\,,\;G\,,\;G\,\cos I\,,\;\theta\,,\;\psi\;$.

~\\

\noindent {\large{\textbf{A.3.~~~Taking variations of $S$}}}\\

 \noindent

 Let us start with the expression (\ref{51}) for the generating function:
 \ba
  S\;=\;-\;t\;T_{kin}\;+\;G\;h\;\cos I\;+\; G\;g\;+\;G\;\int \cos{J}\,\;dl\,+\;C\;\;\;.\;\;\;
  \label{A47}
 \ea
 We see that it depends upon eight variables, some of which are dependent upon others. In brief,
 \ba
  S\;=\;S(\,t\;,\,\;T_{kin}\;,\,\;l\;,\;\;L\,\equiv\,G\;\cos
  J\;,\,\;g\,,\;G\;,\,
        \;h\;,\;\;H\,\equiv\,{G \;\cos I}\,)\;\;\;.~~~~
 \label{A48}
 \ea
 Variation thereof, taken at a fixed time $\;t\;$, will read:
 \ba
 \nonumber
  {\delta}S
     \;=\; \left(\frac{{\partial}S}{{\partial}T_{kin}}\right){\delta}T_{kin}
     + \left(\frac{{\partial}S}{{\partial}h}\right){\delta}h
     + \left(\frac{{\partial}S}{{\partial}(G\; \cos I)}\right){\delta}(G\; \cos
     I)~~~~~~~~~~~~~~~~~~~
  \ea
  \\
  \ba
  \nonumber
     + \left(\frac{{\partial}S}{{\partial}G}\right){\delta}G
     + \left(\frac{{\partial}S}{{\partial}g}\right){\delta}g
     + \left(\frac{{\partial}S}{{\partial}(G\;\cos J)}\right){\delta}(G\;\cos J)
     + \left(\frac{{\partial}S}{{\partial}l}\right){\delta}l
  \ea
  \\
  \ba
 \nonumber
 =\;-\; t\;\delta T_{kin} \;+ \;G\;\cos I\;\delta h \;+\;h\;\delta (G\;\cos I)
 ~~~~~~~~~~~~~~~~~~~~~~~~~~~~~~~\\
  \nonumber \\
  \nonumber\\
     + \left(g + {\int}{\cos J}\; dl\right){\delta}G
      + G\;{\delta}g + G{\int}{\delta}(\cos J)\; dl~~~.~~~
  \label{A51}
 \ea
 Our goal is to simplify (\ref{A51}), having in mind that the variables emerging there are
 not all mutually independent. To that end, let us employ expression (\ref{A3}).
 Its differentiation would yield the expression
 \begin{equation}
 d\;g\;-\;d\;(\phi-h)\;\,\cos I\;=\;-\;\cos{J}\;\,d\left(l\,-\;\psi
 \right)\;+\;\sin{J}\;\,\sin{(l - \psi)}\;d\theta\;\;\;,
 \label{A52}
 \end{equation}
 where the Euler angles $\phi,\,\theta,\,\psi$ determine the orientation
 of the body relative to some inertial reference frame. If, however, we
 perform {\em{variation}} of (\ref{A3}), for a fixed orientation of the
 body relative to the inertial frame, then we shall get simply $\,{\delta}
 g\,=\,-\,\cos J\,{\delta}l\,-\,\cos I\,{\delta}h\,$. Multiplying this by
 $G$, we arrive at
 \begin{equation}
 G\;{\delta}g\;=\;-\;G\;\cos J\;{\delta}l\;-\;G\;\cos I\;{\delta}h\;\;\;,
 \label{A53}
 \end{equation}
 substitution whereof into (\ref{A51}) entails
 \begin{align}
 {\delta}S\;=\;
 & - t\;{\delta}T_{kin} \;+ \;h\;{\delta}(\,G\; \cos I\,)
 \;+\; \left(\,g \;+\; {\int}{\cos J}\; dl\,\right)\;{\delta}G
 \nonumber \\
 & +\;G\;{\int}\,{\delta}(\cos J)\; dl \;-\; G\; \cos J\; {\delta}l\;\;
 \;.
 \label{A54}
 \end{align}
 Next, we shall recall the formula for the Binet Ellipsoid, expressed
 through the directional cosines defined in (\ref{A12}. According to
 (\ref{A10}) and (\ref{A12}), we have:
 \begin{equation}
 \frac{({\alpha}^{'})^2}{I_1} + \frac{({\beta}^{'})^2}{I_2} +
 \frac{({\gamma}^{'})^2}{I_{3}}\; =\; \frac{2\; T_{kin}}{G^2}\;\;\;.
 \label{}
 \end{equation}
 Insertion of the (evident from Fig.~\ref{frames}) relations
 \ba
 \nonumber
 \alpha'~=~\sin J~\sin l~~~,~~~~~~{\beta}^{'}~=~\sin J~\cos l~~~~~~
 {\gamma}^{'}~=~\cos J~~~,
 \ea
 into the above formula will yield
 \begin{equation}
 \sin^2 J\; \left(\frac{\sin^{2}l}{I_{1}} + \frac{\cos^{2}l}{I_{2}}
 \right) + \frac{\cos^{2}J}{I_{3}}\; =\; \frac{2\; T_{kin}}{G^2}\;\;\;,
 \label{}
 \end{equation}
 or, equivalently,
 \begin{equation}
 (1 - ({\gamma}^{'})^2)\; \left(\frac{sin^2l}{I_1} + \frac{cos^2l}{I_2}
 - \frac{1}{I_3}\right) = \frac{2\; T_{kin}}{G^2} -
 \frac{1}{I_3}\;\;\;.
 \label{A55}
 \end{equation}
 Differentiation of the above will lead us to
 \begin{align}
 2\; \frac{dT_{kin}}{G^2}\; - 4\; \frac{T_{kin}\; dG}{G^3}\; =\;
 & -2\; {\gamma}^{'}\; d({\gamma}^{'}) \left(\frac{\sin^2 l}{I_1}
 +\frac{\cos^2 l}{I_2} - \frac{1}{I_3}\right)
 \nonumber \\
 \nonumber\\
 & + 2\; {\alpha}^{'} {\beta}^{'} \left(\frac{1}{I_1}
 -\frac{1}{I_{2}}\right) dl\;\;\;.
 \label{A56}
 \end{align}
 If we now turn the differentials into variations, and multiply both sides
 by $\;G^2\; dt\; /\; 2\;$, we shall obtain:
 \begin{align}
  \left({\delta}T_{kin}\; - 2\; \frac{T_{kin}\; {\delta}G}{G}\right)\; &dt
 \nonumber \\
    &=\; - G^2\; dt\; {\gamma}^{'}\; {\delta}{\gamma}^{'}
     \left(cos^2 l\left(\frac{1}{I_2} - \frac{1}{I_1}\right)
      + \left(\frac{1}{I_1} - \frac{1}{I_3}\right)\right)
 \nonumber \\
 \nonumber\\
    &+ G^2\; dt\; {\alpha}^{'} \; {\beta}^{'} \left(\frac{1}{I_1}
      - \frac{1}{I_2}\right) \delta l\;\;\;.
  \label{A57}
 \end{align}
 Consider the second term on the right hand side of (\ref{A57}). We know from (\ref{A30}) that
 \begin{equation}
  G\; dt\; \alpha' \; {\beta}^{'} \; \left( \frac{1}{I_{1}}
    - \frac{1}{I_{2}} \right) \; = \; - \;d {\gamma}^{'} \;\;\;,
  \label{A58}
 \end{equation}
 so we can write the second term on the right hand side of (\ref{A57}) as
  \begin{equation}
  G^{2}\; dt\; {\alpha}^{'} \; {\beta}^{'} \left(\frac{1}{I_1}
    - \frac{1}{I_2}\right) {\delta}l \;=\; -\;G\; d {\gamma}^{'} \; {\delta}l\;\;\;.
  \label{A59}
  \end{equation}
The first term on the right hand side of (\ref{A57}) can be
written down as
  \begin{align}
  - \;G^2\; &dt\; {\gamma}^{'}\; \delta {\gamma}^{'}
    \left[ \cos^2 l\; \left(\frac{1}{I_2} - \frac{1}{I_1}\right)
  + \left(\frac{1}{I_1} - \frac{1}{I_3}\right)\right]
 \nonumber \\
 \nonumber\\
  &=\; - \;G^2\; dt\; {\gamma}^{'}\; \delta {\gamma}^{'}
    \left[ \frac{({\beta}^{'})^2}{1 - ({\gamma}^{'})^2}\; \left(\frac{1}{I_2}
    - \frac{1}{I_1}\right) + \left(\frac{1}{I_1} -
    \frac{1}{I_3}\right)\right]
 \nonumber \\
 \nonumber\\
  &=\; - G\; {\gamma}^{'}\; \delta {\gamma}^{'}
    \left(
       \frac{({\beta}^{'})^2}
         {1 - ({\gamma}^{'})^2}\; \frac{d{\gamma}^{'}}{{\alpha}^{'}\; {\beta}^{'}} +
         \frac{d{\beta}^{'}}{{\alpha}^{'}\; {\gamma}^{'}}
    \right)
    \nonumber \\
    \nonumber \\
  &=\; \frac{- G\; {\delta}{\gamma}^{'}}{{\alpha}^{'}}\;
    \left(
       \frac{{\beta}^{'}\; {\gamma}^{'}}{1 - ({\gamma}^{'})^{2}}\; d {\gamma}^{'}
       + d{\beta}^{'}
    \right)
    \nonumber \\
    \nonumber \\
  &=\; \frac{- G\; \delta {\gamma}^{'}}{{\alpha}^{'} (1 - ({\gamma}^{'})^{2})}\;
    \left(
       {\beta}^{'}\; {\gamma}^{'} \; d{\gamma}^{'}
       + (1 - ({\gamma}^{'})^{2})\; d{\beta}^{'}
    \right) \;\;\;.
  \label{A60}
  \end{align}
We obtained the first equality by noting that the definitions of
the direction cosines yield
 \begin{equation}
  \cos^{2}\; l\; =\; \frac{({\beta}^{'})^2}{({\alpha}^{'})^2
    + ({\beta}^{'})^2}\; =\; \frac{({\beta}^{'})^2}{1 - ({\gamma}^{'})^2}\;\;\;,
  \label{A61}
 \end{equation}
 while the second equality comes from the equations of motion (\ref{A28} - \ref{A30}) in Appendix A.2.
 Finally, when we look at $\;d{\beta}^{'}\;$, we see that
  \begin{align}
  d{\beta}^{'} \; &=\; d(\sin\; J\; \cos\; l)
    \;=\; d(\sqrt{1 - ({\gamma}^{'})^{2}}\; \cos\; l)
 \nonumber \\
 \nonumber \\
    &=\, -\frac{{\gamma}^{'}\, d{\gamma}^{'}\; \cos l}{\sqrt{1
     - ({\gamma}^{'})^2}} - \sqrt{1 - ({\gamma}^{'})^2}\; \sin l\, dl
    \,=\, -\frac{{\gamma}^{'}\, {\beta}^{'}\, d{\gamma}^{'}}{1
     - ({\gamma}^{'})^2} - {\alpha}^{'}\, dl\;\;.
  \label{A62}
  \end{align}
We can now write (\ref{A60}) as
  \begin{align}
  &\frac{- \;G\; {\delta}{\gamma}^{'}}{{\alpha}^{'}(1
    - ({\gamma}^{'})^2)}\; \left({\beta}^{'}\; {\gamma}^{'}\; d{\gamma}^{'}
    + (1 - ({\gamma}^{'})^2)\; d{\beta}^{'}\right)
 \nonumber \\
  &= \frac{- G\; {\delta}{\gamma}^{'}}{{\alpha}^{'}(1
    - ({\gamma}^{'})^2)}\; \left({\beta}^{'}\; {\gamma}^{'} \; d{\gamma}^{'}
    + (1 - ({\gamma}^{'})^2)\; \left( -\frac{{\gamma}^{'}\; {\beta}^{'}\;
    d{\gamma}^{'}}{1 - ({\gamma}^{'})^2} - {\alpha}^{'}\; dl\right)\right)
 \nonumber \\
  &=\; G\; dl\; {\delta}{\gamma}^{'}
 \nonumber \\
  &=\, G\, {\delta}(\cos J)\, dl\;\;.
  \label{A63}
  \end{align}
 This enables us to write (\ref{A57}) as
  \begin{equation}
  \left({\delta}T_{kin}\; - 2\; \frac{T_{kin}\; {\delta}G}{G}\right) \; dt =
     G\; {\delta}(\cos J)\; dl - G\; d(\cos J)\; {\delta}l\;\;\;.
  \label{A64}
  \end{equation}
Integration of this yields
  \begin{align}
  G{\int}{\delta}(\cos J)\; dl\;
   &=\; \left({\delta}T_{kin}\; - 2\; \frac{T_{kin}\; {\delta}G}{G}\right)\; {\int}dt
   + G\; {\int}{\delta}l\; d{\gamma}^{'}
 \nonumber \\
 \nonumber\\
   &=\; \left({\delta}T_{kin}\; - 2\; \frac{T_{kin}\; {\delta}G}{G}\right)\;
     \frac{u}{n} \;+ \;G\; \cos J\; {\delta}l\;\;\;.
  \label{A65}
  \end{align}
Insertion of this result into (\ref{A54}) brings up the following
 \ba
 {\delta}S\,=\,\left(\,\frac{u}{n} \,- \,t\,\right)\;{\delta}T_{kin}\,+\,h\;
 {\delta}(G\;\cos I)\,+\,\left(g\,+{\int}{\,\cos J}\; dl\,-\,2\,\frac{T_{kin}}{G}
 \,\frac{u}{n}\right)\, {\delta}G\;\;.~~~~
  \label{A66}
 \ea
\paragraph{}
 Finally, getting rid of $\;\delta G\,\left(\,\int \cos J\,
 dl\,-\,2\,({T_{kin}}/{G})\,({u}/{n})\right)\;$ by means of
(\ref{54}), we arrive at the desired formula (\ref{60}).

~\\

\noindent {\large{\textbf{A.4.~~~Proof of formula (\ref{52})}}}\\

 Since we are now talking about unperturbed rotation, the invariable
 plane is inertial and we can employ our formulae (\ref{5} - \ref{7})
 for the body-frame-related angular velocity,
 with angles $\,g\,,\;J\,,\;l\,$ inserted instead of $\,\phi\,,\;\theta\,,\;\psi\,$:
 \begin{eqnarray}
  \omega_1 &=& \dot{g}\;\sin J\;
  \sin l\;+\;\dot{J}\;\cos l\;\;\;,
   \label{omega1}
   \label{9995}\\
  \omega_2 &=&
  \dot{g}\;\sin J\;\cos l\;-\;\dot{J}\;\sin l\;\;\;,
   \label{omega2}
   \label{9996}\\
  \omega_3 &=& \dot{l}\;+\;\dot{g}\;\cos J\;\;\;.
   \label{omega3}
   \label{9997}
 \end{eqnarray}
 Therefrom we extract two derivatives:
 \[
 \begin{array}{l}
 \dot{g} = \Frac{1}{\sin J}\, (\omega_1  \sin l + \omega_2  \cos l)
 \;\;\;,\\
 \nonumber\\
 \dot{l}\, \cos J = \omega_3 \,\cos J - \Frac{\cos^2
 \theta}{\sin J}(\omega_1  \sin l + \omega_2  \cos l) ,
 \end{array}
 \]
summation whereof yields
 \[
 \begin{array}{l}
 \dot{g} + \dot{l}\, \cos J  =
   \omega_3 \,\cos J + \Frac{\textstyle{1 - \cos^2 J}}{\sin J}(\omega_1  \sin l +
   \omega_2
   \cos l) =
    \omega_3 \,\cos J + \sin J\, (\omega_1  \sin l + \omega_2  \cos
    l)\;\;.\;\;\;
 \end{array}
 \]
 This expression, in combination with
 \ba
 \nonumber
 I_1\,\omega_1\;=\;G\;\sin J \; \sin l  \;\;\;,\;\;\;\;\;
 I_2\,\omega_2\;=\;G\;\sin J \; \cos l  \;\;\;,\;\;\;\;\;
 I_3\,\omega_3\;=\;G\;\cos J \;\;\;,\;\;\;\;\;
 \ea
 entails (\ref{52}).
\\

\noindent {\large{\textbf{A.5.~~~Andoyer's Method}}}\\

 Let $(x_i\,,\, y_i)$ $\,i=1,\ldots,n\,,$ be a set of canonical variables,
 and let the Hamiltonian be some ${\mathcal H}(x_i\,, \,y_i)$, so that
 \begin{equation}
 \label{eqHam}
 \frac{d x_i}{d t} = \frac{\partial \mathcal{H}}{\partial y_i}\;\;, \qquad \Frac{d
 y_i}{d t}= \;- \;\Frac{\partial \mathcal{H}}{\partial x_i}\;\;\;.
 \end{equation}

 Let $z_1, z_2, \ldots z_{2n}$ be some new, not necessarily canonical variables. The
 question raised by Andoyer was: How will the equations of motion look in these new
 variables?

 Making use of the Lagrange brackets,
\[
[u,v] = \sum_i \left(\Frac{\partial x_i}{\partial u}
   \Frac{\partial y_i}{\partial v} - \Frac{\partial y_i}{\partial u}
   \Frac{\partial x_i}{\partial v}   \right),
\]
 and taking into account (\ref{eqHam}), we can write after some
 algebra:\footnote{Note that \[
\Frac{\partial \mathcal{H}}{\partial z_k} = \sum_i \left(
\Frac{\partial \mathcal{H}}{\partial x_i} \Frac{\partial
x_i}{\partial z_k} + \Frac{\partial \mathcal{H}}{\partial y_i}
\Frac{\partial y_i}{\partial z_k} \right),
\] and hence, based on (\ref{eqHam}), \[
\Frac{\partial \mathcal{H}}{\partial z_k} + \sum_i \left(
 \Frac{\partial x_i}{\partial z_k} \Frac{d y_i}{dt}-
 \Frac{\partial y_i}{\partial z_k} \Frac{d x_i}{dt}
\right) = 0.
\]}
\begin{equation}\label{linear}
 \Frac{\partial \mathcal{H}}{\partial z_k} + [z_k,t] + \sum_l [z_k, z_l]
   \Frac{d z_l}{d t} = 0\;\;\;,
\end{equation}
 a linear system wherefrom the derivatives $\,d z_l/d t\,$ can be obtained.

 At this point, Andoyer introduces a new function $J_u$ given as
 \begin{equation}
 \label{defJ}
 J_u =  \Frac{\partial K}{\partial u} +
 \sum_i y_i  \Frac{\partial x_i}{\partial u},
 \end{equation}
 where $K$ is an arbitrary function (later on it will be set zero).

 Obviously,
 \ba
 [u, v] =  \Frac{\partial J_u}{\partial v}
 - \Frac{\partial J_v}{\partial u}.
 \ea
 The linear system (\ref{linear}) may be converted into
 \ba
 \Frac{\partial \mathcal{H}'}{\partial z_k} + \Frac{\partial J_{z_k}}{\partial t}+
 \sum_l \left(\Frac{\partial J_{z_k}}{\partial z_l} -
 \Frac{\partial J_{z_l}}{\partial z_k} \right)
 \Frac{d z_l}{d t} = 0\;\;\;,
 \ea
 with $\;\mathcal{H}'\,= \,\mathcal{H}\,-\,t\,$.

Let us assume now the following hypothesis: the $2n$ variables $z_k$
are split into two parts $p_j$ and $q_j$, ($j=1,\ldots n$), in such
a way that
 \begin{equation}
 \label{conditionJ}
 J_{q_j} = 0, \quad \mbox{and} \quad J_{p_j}  = J_{p_j} (q_1,\ldots,
 q_n; t).
 \end{equation}
Hence, the previous system is split into
 \ba
 \Frac{\partial \mathcal{H}'}{\partial p_k} + \Frac{\partial J_{p_k}}{\partial t}
 + \sum_j \Frac{\partial J_{p_k}}{\partial q_j} \Frac{d q_j}{d t} =
 0,\quad \Frac{\partial \mathcal{H}'}{\partial q_k} - \sum_j \Frac{\partial
 J_{p_j}}{\partial q_k}
 \Frac{d p_j}{d t} = 0.
 \ea

 Let us now define $q^*_j = J_{p_j}$, and replace $q_j$ with these new
 variables. Since  $q_j$ are functions of $q^*_j$ and $t$, we get:
 \ba
 \Frac{\partial \mathcal{H}'}{\partial q_k} = \sum \Frac{\partial \mathcal{H}'}{\partial
 q^*_k}\Frac{\partial q^*_j}{\partial q_k}\;\;\;,
 \ea
 whence the previous equations acquire the form of
 \ba
 \Frac{d p_j}{d t} = \Frac{\partial \mathcal{H}'}{\partial q^*_j},\qquad
 \Frac{d q^*_j}{d t} = -\Frac{\partial \mathcal{H}'}{\partial p_j}\;\;\;.
 \ea
 This shows that $p_j$ are the conjugate momenta of $q^*_j$.

Let us now consider a rigid-body motion. Let $\phi,\theta, \psi$ be the Euler angles and
$\Phi,\Theta, \Psi$ be their conjugate moments. If in the definition (\ref{defJ}) we set
the function $K$ to be zero, we shall obtain:
 \ba
 J_\alpha =  \Phi \Frac{\partial \phi}{\partial \alpha} +
   \Theta \Frac{\partial \theta}{\partial \alpha} +
   \Psi \Frac{\partial \psi}{\partial \alpha},
 \ea
where $\alpha$ denotes any Euler angle. By definition, a conjugate momentum is a partial
derivative of the kinetic energy with respect to the time derivative of the
corresponding Euler angle. Thus
 \ba
 J_\alpha =  \Frac{\partial \phi}{\partial \alpha}
 \Frac{\partial T}{\partial \dot\phi } +
   \Frac{\partial \theta}{\partial \alpha}
     \Frac{\partial T}{\partial\dot\theta } +
   \Frac{\partial \psi}{\partial \alpha}
   \Frac{\partial T}{\partial \dot\psi }\;\;\;,
 \ea
which we can be re-written as
 \ba
 J_\alpha =  \dot\phi_\alpha \Frac{\partial T}{\partial \dot\phi } +
   \dot\theta_\alpha
     \Frac{\partial T}{\partial\dot\theta} +
   \dot\psi_\alpha   \Frac{\partial T}{\partial \dot\psi},
 \ea
where $\dot\phi_\alpha,\dot\theta_\alpha,\dot\psi_\alpha$ are the
virtual angular velocities with respect to $\alpha$; the virtual
motion is an instantaneous rotation about the origin, at an angular
velocity ${\omegabold}_\alpha = \,(\,\omega_{1\alpha}\,,\,
\omega_{2\alpha}\,,\, \omega_{3\alpha}\,)\,$. Thus, the relation
between ${\omegabold}_\alpha $ and the derivatives
$\dot\phi_\alpha,\dot\theta_\alpha,\dot\psi_\alpha$, is the same as
the relation between the angular-velocity vector ${\omegabold}$ and
the derivatives $\dot\phi,\dot\theta,\dot\psi$.

The kinetic energy is a quadratic form in the derivatives
$\dot\phi,\dot\theta,\dot\psi$. So the above expression shows that
$\,J_\alpha\,$ is the polar form of the quadratic form $\,T\,$ with
respect to $\dot\phi_\alpha,\dot\theta_\alpha,\dot\psi_\alpha$.
Thus, since
 \ba
  2 T = I_1
\omega_1^2 +  I_2 \omega_2^2 +  I_2 \omega_3^2 =\bf{G} \cdot {\omegabold}\;\;\;,
 \ea
  we have:
 \ba
 J_\alpha = I_1 \omega_1 \omega_{1\alpha} +  I_2 \omega_2\omega_{2\alpha}  +  I_2
\omega_3\omega_{3\alpha} = \bf{G} \cdot {\omegabold}_\alpha \;\;\;.
 \ea

Having arrived to this point, Andoyer introduced the invariable plane perpendicular to
the angular momentum vector $\bf{G}$, and computed the quantities $J_\alpha$ in terms of
the six variables $\,l\,,\,g\,,\,h\,,\,I\,,\,G\,,\,J\,$ defined as on Fig.~3.

Firstly, $\,J_G \,=\, 0\,$ because $\,\phi\,,\, \theta\,,\, \psi$ do not depend on
$\,G\,$. Next, as vectors $\,\omegabold_I\,$ and $\,\omegabold_J\,$ point in the
directions of the nodes on the invariable plane, then $\,J_I \,= \,{\bf{G}}\, \cdot\,
{\omegabold}_I \, =\, 0\,$ and $\,J_J \,= \,{\bf{G}}\, \cdot\,{\omegabold}_J\, =\, 0\,$.
On the other hand, $\,J_g \,= \,{\bf{G}} \,\cdot\, {\omegabold}_g \, =\, G\,$, and
$\,J_h \,= \,{\bf{G}}\, \cdot\, {\omegabold}_h \,=\, G\, \cos I\,$, and also $\,J_l\,
=\, {\bf{G}}\, \cdot\, {\omegabold}_J\, =\, G \cos J\,$.

Thus, to satisfy the conditions (\ref{conditionJ}) and to get a canonical set of
variables, we must choose the new conjugate momenta as
 \ba
 L = J_l = G \cos J, \quad G = J_g\, , \quad H = J_h = G \cos I.
 \ea

\section*{Acknowledgments}

The authors are grateful to Jerry Marsden for reading this text and
making valuable comments.

 \newpage

\end{document}